\documentclass{article}

\usepackage{arxiv}

\usepackage[utf8]{inputenc} 
\usepackage[T1]{fontenc}    
\usepackage{hyperref}       
\usepackage{url}            
\usepackage{booktabs}       
\usepackage{amsfonts}       
\usepackage{nicefrac}       
\usepackage{microtype}      
\usepackage{lipsum}
\usepackage{graphicx}
\graphicspath{ {./images/} }

\usepackage{amsmath}
\usepackage[capitalise]{cleveref}			
\usepackage{algorithm}	
\usepackage{algorithmic}
\usepackage{enumitem}
\usepackage{siunitx}
\usepackage{multirow}

\title{Parameter identification for a damage
model using a physics informed neural
network}

\author{
 Carlos J. G. Rojas \\
  School of Mechanical Engineering\\
  University of Campinas \\
  \texttt{c212011@dac.unicamp.br} 
   \And
 Marco L. Bitterncourt \\
 School of Mechanical Engineering\\
 University of Campinas \\
 \texttt{mlb@fem.unicamp.br}
  \And
 José L. Boldrini \\
   School of Mechanical Engineering\\
 University of Campinas \\
  \texttt{josephbold@gmail.com} 

}

\begin{document}
\maketitle
\begin{abstract}
This work applies concepts of artificial neural networks to identify the parameters of a mathematical model based on phase fields for damage and fracture. Damage mechanics is the part of
the continuum mechanics that models the effects of micro-defect formation using state variables
at the macroscopic level. The equations that define the model are derived from fundamental
laws of physics and provide important relationships between state variables. Simulations using
the model considered in this work produce good qualitative and quantitative results, but many
parameters must be adjusted to reproduce a certain material behavior. The identification of model
parameters is considered by solving an inverse problem that uses pseudo-experimental data to
find the values that produce the best fit to the data. We apply a physics informed neural network
and combine some classical estimation methods to identify the material parameters that appear
in the damage equation of the model. Our strategy consists of a neural network that acts as an
approximating function of the damage evolution with its output regularized using the residue
of the differential equation. Three stages of optimization seek the best possible values for the
neural network and the material parameters. The training alternates between the fitting of only
the pseudo-experimental data or the total loss that includes the regularizing terms. We test the
robustness of the method to noisy data and its generalization capabilities using a simple physical
case for the damage model. This procedure deals better with noisy data in comparison with a
PDE-constrained optimization method, and it also provides good approximations of the material
parameters and the evolution of damage.
\end{abstract}


\section{Introduction}

 Parameter estimation is an important step in the development of accurate models on natural sciences, physics, engineering, and many other disciplines. Mathematical models developed to approximate dynamic processes often involve differential equations with unknown parameters. In a parameter identification problem, a metric that measures the error is minimized to fit a model prediction with observed data.  As described in \cite{Mehrkanoon2012}, there are two main categories of methods used to estimate the parameters in the governing equation of a model. In the first category of methods, the differential equations are solved adopting random initial values for the parameters and their predictions are compared with experimental data. An objective function is defined to quantify the difference between the expected and obtained results and the model parameters are updated applying an optimization procedure. According to \cite {Moles2003}, this approach requires a high computational cost and almost $90 \%$ of the computation time is required to solve the model equations.The second category of methods substitute the solution of the governing equations adopting a functional approximation. Using this approximation, the required derivatives are calculated and the residue of the differential equation is constructed. Subsequently, this residue is minimized by adopting an optimization algorithm and the model parameters are estimated \cite{Mehrkanoon2012}.

 A novel methodology for parameter estimation with the introduction of artificial neural networks as model approximators was introduced in \cite{Dua2011}. This work is part of the second category presented in the last paragraph and can be defined as a decomposition algorithm \cite{Dua2011b}. The author divides the problem into two steps, first an artificial neural network approximates the model after a process of training with measured data, and then the residue of the differential equation is defined through derivatives of the neural network.  The parameters are part of the expression defining the residue and estimated employing an optimization procedure. Most of the examples presented in his article are from models of chemical reactions described by systems of ordinary differential equations. In a subsequent work, the ideas described in \cite{Dua2011} were generalized. A simultaneous approach  is introduced  in \cite{Dua2011b}, where the solution of ordinary differential equations and determination of the model parameters is performed at the same time. The objective function of the optimization process for this methodology is composed of a term that relates the differences between the predictions and observed data and the residue to satisfy the governing equations of the model. The method showed good results for different examples solved with only one hidden layer and a small number of nodes.

More recently, a data-driven discovery algorithm for estimation of parameters in partial differential equations was proposed in  \cite{raissi2017physics}. Their methodology is called physics informed neural networks (PINNs) and the main differences with the work of \cite{Dua2011b} is the possibility to include boundary and initial conditions in the loss function of the neural network and the adoption of automatic differentiation to compute the derivatives of the network. These physics informed neural networks (PINNs) were applied in benchmark problems using a relatively small amount of data (system’s solutions) and regularizing the system with the physics laws represented by differential equations. Similarly, in \cite{alex2018learning}, a PINN was applied to approximate the space-dependent coefficient in a linear diffusion equation and the constitutive relationship in a non-linear diffusion equation. Other successful applications can be found in  \cite{raissi2018multistep,raissi2018hidden,Raissi2019,Tartakovsky2019,Tipireddy2019,Meng2020}, where the inclusion of differential equations or constitutive equations as part of the loss function in neural networks have demonstrated to be efficient, accurate and suggest great promise for future applications.

There are also versatile data-driven approaches where even the differential operators of the models are estimated from data. In \cite{long2017pdenet}, the authors presented a method where the differential and nonlinear operators of a governing equation are learned without the definition of a fixed equation. The use of a feedforward neural network, called PDE-net, allowed the discovery of a hidden model using observational data and was also able to predict its dynamical behavior.  Some examples using convection-diffusion equations uncovered the hidden equations from simulated data and provided good approximations for the dynamic behavior. Subsequently, in \cite{Rudy2019}, a general method to identify the governing equations of a given model was proposed. This work is called a  PDE-FIND data-driven model and the terms of the governing equations are selected from a library with linear, nonlinear, time and space differential operators. The method was tested in the identification of four canonical models and produced accurate approximations.

The starting point for the parameter estimation performed in this article is the work developed in \cite{raissi2017physics}. We apply their methodology to identify the material parameters of the damage equation considering that the differential operators were derived with mathematical consistency following the basic principles of continuum mechanics. In addition, we consider that the parameters of the model are identifiable and that the neural network approximation is suitable in the case of ill-posed problems. 

The identification is based on the construction of an objective function that measures  differences between pseudo-experimental data and results obtained using a neural network.
Though few consistent mathematical theories explain the suitability of neural networks for inverse problems, many applications have achieved good results even for ill-posed problems \cite{adler2017, seo_2019,li_nett:_2020}. We formulate an optimization problem where the minimization of an objective function leads to optimal parameters that reproduce the principles of a physics model. Instead of using costly approaches, such as the computation of forward solutions for each optimization step, we propose here the use of physics informed neural networks. With this method, the solution is approximated using artificial neural networks and the parameters are estimated from the residue of a partial differential equation.

\section{Parameter identification in differential equations} \label{pid_sec}

Solution of the governing equations of physics models has been extensively studied in mathematics and applied in engineering fields. However, to obtain realistic results from the solution of governing equations, a model has to be validated and calibrated using experimental data. The process of validation and calibration requires the solution of an inverse input-output mapping and it is necessary to find the causes that lead to that state \cite{Buljak2012}.

A parameter identification in differential equations is a class of inverse problem where the unknown inputs of a model are the parameters entering into the governing equations.  Identification problems arise in fields such as geophysics, fluids, structural mechanics, electromagnetics, biomedical  and thermal sciences and there has been a steady interest in developing efficient estimation approaches for different applications.

As previously introduced, an important step in the formulation of a model is the connection of its governing equations with experimental data that comes from observations. A general mathematical form used to represent  several phenomena is
\vspace{5mm}
\begin{equation}
\mathcal{F}(x_1, ..., x_j, u, \frac{\partial u}{\partial x_1},...,\frac{\partial u}{\partial x_j},...,\frac{\partial^2 u}{\partial x_1 \partial x_j} ,  \lambda ) = 0,
\vspace{5mm}
\label{Eq1}
\end{equation}
\noindent where $u(x)$ is the state variable, $x$ are the input variables  $(x_1,...,x_j)^T$ and $\lambda$ is the vector of parameters $(\lambda_1,...,\lambda_k)^T$ . These parameters can be values defining a physical entity directly or coefficients in relationships that describe a physical process  \cite{parameter2019,frasso_parameter_2016}.

The data collected from experimental observations is represented by
\vspace{5mm}
\begin{equation}
\hat{u}(x)  = u(x) + \epsilon
\label{Eq2}
\vspace{5mm}
\end{equation}
\noindent where $\epsilon$ is assumed to be an independent normally distributed measurement error with mean zero and constant variance.

The objective of parameter identification in a model described by \cref{Eq1} is to find a set of parameter estimates $\hat{\lambda}$ , that leads to minimal differences between the observed data $\hat{u}(x)$ and the solution of the differential equation  $u(x,\hat{\lambda})$.
Some important issues that need to be considered in this type of problem are the model suitability to represent the experimental data and if it is possible to uniquely identify the parameters with the observations available. As presented in \cite{JADAMBA2011},  inverse problems are frequently  ill-posed  in the sense of Hadamard. Some of the causes of ill-conditioning are an insufficient approximation of  models, data affected with noise and the lack of additional constraints. 

The methods to estimate parameters in differential equations are divided into two major categories. The first category groups deterministic approaches where it is assumed that the state variable of the model is completely defined by its parameters, boundary conditions, and initial conditions. In this case, possible disturbances that can arise in the mathematical formulation of the model, the approximate solution of the governing equations and the observations are not directly accounted for in the estimation of the parameters. Conversely, in the second category, the use of stochastic methods systematically includes these uncertainties using frequentist or Bayesian approaches \cite{varziri_parameter_2008}. 

Several methods to estimate parameters in models described by ordinary or partial differential equations (ODEs or PDEs) have been developed through years \cite{frasso_parameter_2016,varziri_parameter_2008,muller_parameter_2004,ramsay_parameter_2007,cao_penalized_2012,xun_parameter_2013}
. Despite their different degrees of complexity, methods initially developed for identification of parameters in ODEs have been adapted and applied for models described by PDEs.

The first step common in any estimation method is the selection of the criterion to fit the data. A prominent criterion in inverse problems is the least square method which is adequate in cases where the uncertainties can be modeled using Gaussian distributions. This is the preferred standard because it simplifies the calculations in the optimization process but has difficulties dealing with outliers in the data. An alternative to avoid that problem is the least absolute value criterion which reduces the sensitivity to errors introduced by unlikely observations \cite{Tarantola2005}. After the selection of fitting criteria, the next step is to pose the identification of parameters as an optimization problem. The first option is to propose the constrained optimization presented in the following expression:
\vspace{5mm}
\begin{equation}
\begin{aligned}
 \min_{\lambda} \quad & \sum_{i=1}^{N}{(u(x) -\hat{u}(x))^2}\\
\textrm{s.t.} \quad &\mathcal{F}(x_1, ..., x_j, u, \frac{\partial u}{\partial x_1},...,\frac{\partial u}{\partial x_j},...,\frac{\partial^2 u}{\partial x_1 \partial x_j} ,  \lambda ) = 0,
\end{aligned}
\label{Eq3}
\vspace{5mm}
\end{equation}
\noindent where the objective is to fit the state variable of the model ${u}(x)$ to the experimental data $\hat{u}(x)$ using a least square criterion and with the differential equation as the constraint. In many situations governing equations cannot be solved analytically, so, in addition to the data fitting procedure, it is necessary to introduce a numerical method to approximate the solution of the model equations and the sensitivity of the state variable to the parameters of the model. This increases the computational complexity of the problem and requires the implementation of parallel optimized low-level code and high-level abstraction frameworks for automatic differentiation that only some software libraries such as PETSc \cite{petsc-web-page}, DOpElib \cite{dopelib:_2017} and dolfin-adjoint \cite{funke2013framework} have developed.

A second option, widely implemented in identification problems, is the use of function basis expansions to approximate the output of the model. This approach was introduced in the work of \cite{varah_spline_1982}  and its main advantage is the decrease in the computational cost of directly solving the differential equation. Furthermore, it also reduces the propagation of errors and prevents some stability issues. The non-parametric approximation using a linear combination of basis functions $\phi(x)$ is given by
\vspace{5mm}
\begin{equation}
\tilde{u}(x) = \sum_{i=1}^{K} \phi_i(x) c_i,
\label{Eq4}
\vspace{5mm}
\end{equation}
\noindent where $c_i$ are the basis coefficients and $K$ is the number of collocation points necessary to fit the data.

\subsection{Methods to estimate parameters using function approximations}
The simplest procedure to estimate parameters using basis expansion is known as the two-step method \cite{Dua2011,varah_spline_1982}. In this methodology, the coefficients of the base $c_i$ are approximated considering only the fitting criteria
\vspace{5mm}
\begin{equation}
\begin{aligned}
 \min_{c_i} \quad & \sum_{i=1}^{N}{(\tilde{u}(x)-\hat{u}(x))^2},
 \end{aligned}
\label{Eq5}
\vspace{5mm}
\end{equation}
\noindent and then in a separated stage, the parameters $\lambda$ are estimated using the residue of the differential equation evaluated in $N$ collocation points as
\vspace{5mm}
\begin{equation}
\begin{aligned}
 \min_{\lambda} \quad & \sum_{i=1}^{N}{ (\mathcal{F}(x_1, ..., x_j, \tilde{u}, \frac{\partial \tilde{u}}{\partial x_1},...,\frac{\partial \tilde{u}}{\partial x_j},...,\frac{\partial^2 \tilde{u}}{\partial x_1 \partial x_j} ,  \lambda ))^2},
 \end{aligned}
\label{Eq6}
\vspace{5mm}
\end{equation}
\noindent where the evaluation of the residue in collocation points is equivalent to use a Monte Carlo integration method  and only the values of $\lambda$ are approximated maintaining fixed the coefficients of the basis expansion \cite{remco_delft}.

One difficulty of this procedure is that the function has to capture the behavior of the system without including the noise and uncertainties present in the observations. This issue can be solved by refining the number and position of the collocation points or including a penalty term in \cref{Eq5} to set a balance between overfitting and underfitting of the experimental data. The penalty term can be a high order derivative or, in methods such as principal differential analysis \cite{poyton_parameter_2006}, the residue of the differential equation  as expressed in the following equation
\vspace{5mm}
\begin{equation}
\begin{aligned}
 \min_{c_i} \quad & \sum_{i=1}^{N}{(\tilde{u}(x)-\hat{u}(x))^2} + \alpha_r \sum_{i=1}^{N}{ (\mathcal{F}(x_1, ..., x_j, \tilde{u}, \frac{\partial \tilde{u}}{\partial x_1},...,\frac{\partial \tilde{u}}{\partial x_j},...,\frac{\partial^2 \tilde{u}}{\partial x_1 \partial x_j} ,  \lambda ))^2}, \\
\end{aligned}
\vspace{5mm}
\label{Eq7}
\end{equation}
\noindent where the weight $\alpha_r$ controls the amount of regularization and the residue of the model is computed using initial estimates of the parameters $\lambda$. In this case $\alpha_r$ can be also interpreted as an smoothing parameter that manages the fidelity of the approximation to the model \cite{zhang_estimating_2017}.
In \cite{poyton_parameter_2006}, the authors developed a refined principal analysis, a method that employs \cref{Eq7} to estimate the coefficients of the basis functions and also the parameters of the model. This procedure is executed iteratively between the estimate of the coefficients and the parameters of the model until their estimates converge.

The work in \cite{ramsay_parameter_2007} provides important ideas to be considered in the development of other approximation strategies. They divide the variables estimated into two classes, one for the parameters of the model and the other for the coefficients of the basis function. This distinction is important because the optimization problem is directly concerned with the parameters of the model which the authors called as structural for their main importance. On the other hand, the coefficients of the function basis are designated as nuisance parameters due to their secondary role in the overall identification of the model. Another relevant difference between these two classes is that the number of nuisance parameters exceeds the structural parameters by a significant amount. The last argument is one of the reasons for the authors to avoid the estimation of both parameters at the same time. They use two different levels of optimization, in a similar way as explained before. In the inner level, the optimization only searches for better nuisance parameters and in the outer level, the structural parameters are refined using a different criterion. A third level in the optimization is possible to find the best value of the weight in the penalized term but the authors adjusted it using some heuristics \cite{ cao_parameter_2007}.

Finally, we close our revision of the methods to estimate parameters in differential equations mentioning that the function basis expansion can be substituted for any function approximation procedure desired. In this work, we are interested in the application of neural networks as an approximator of the state variables of a model. The work presented in \cite{Dua2011} was one of the first to introduce artificial neural networks using the two-step estimation approach and in \cite{Dua2011b,raissi2017physics}  were implemented simultaneous searches of structural and nuisance parameters in models described by ODEs and PDEs, respectively. In the following sections, we present an estimation methodology that combines some of the ideas presented in this subsection and employs the general structure proposed in \cite{raissi2017physics}.



\section{Damage model} \label{damage model}

Damage mechanics is a part of solid mechanics that allows a better understanding of the deterioration of materials and tries to predict its implication for mechanical integrity. Although damage involves the creation of microvoids and microcracks, discontinuities at a large scale of the medium, it has been introduced as a continuous variable that represents these volume and surface defects \cite{LemaitreDesmorat}.

Different strategies have been used in the development of models of damage, fracture and fatigue  in elastic solids but few works have followed thermodynamically consistent frameworks. This study uses the damage and fatigue model developed in  \cite{boldrini2016non}, which addresses some mathematical deficiencies that have not been considered in previous works.

In \cite{boldrini2016non}, it is proposed a general thermodynamically consistent non-isothermal continuum framework for the evolution of damage, fatigue and fracture in materials under the hypothesis of small deformation. The approach followed  is based on the use of conservation of mass, the principle of virtual power (PVP) and the first and second law of thermodynamics. In addition to the classical principles, it uses the phase field methodology to introduce fatigue and damage behavior. The kinematic descriptor for damage is a dynamic variable and its evolution is obtained from the PVP. The constitutive relations that define the governing equations of the model are expressed in terms of the free energy potential and the associated pseudopotential of dissipation for any given material  \cite{boldrini2016non,haveroth2018comparison}.

The mentioned general framework is described below for a one-dimensional domain $\Omega=[a,b]$, linear elastic isotropic material, displacements only in the axial direction $x$, isothermal  case and including only the effects of damage in the model.
In such a situation, the model developed in \cite{boldrini2016non} consists of a coupled system of dynamic equations with the evolution of displacement  $u$  
and damage phase field $\varphi$ given by
\vspace{5mm}
\begin{equation}
\rho \frac{\partial^2 u }{\partial t^2 }=\frac{\partial }{\partial x }\Bigl((1-\varphi)^{2} E\frac{\partial u}{\partial x }\Bigr) + f_r(x),\\
\label{Eq8}
\end{equation}
\begin{equation}
\lambda_c\frac{\partial \varphi }{\partial t }=\frac{\partial }{\partial x }\Bigl( g_c\gamma_c\frac{\partial\varphi  }{\partial x } \Bigr)+(1-\varphi)E\Bigl(\frac{\partial u}{\partial x } \Bigr)^{2}-g_c\frac{\varphi}{\gamma_c}.\\
\label{Eq9}
\vspace{5mm}
\end{equation}
\noindent In these equations, $g_c$ is the Griffith fracture energy, $\gamma_c$ is a parameter associated to the width of the damage phase field layers, $\lambda_c$ is related to the rate of damage change, $E$ represents the Young's modulus and $f_r(x)$ is body load function.  \cref{Eq8} describes the evolution of displacement $u$ and \cref{Eq9} is the governing equation of the phase field damage. The variable $\varphi$  represents the volumetric fraction of damaged material such that  $\varphi$  = 0 for virgin material,  $\varphi$  = 1 for fractured material and 0 $<$ $\varphi$  $<$ 1 for damaged material.

There are many possibilities for the boundary conditions of the governing equations. For the first equation, the displacement or the stress are given on the boundary. In the damage equation, the standard boundary condition is a homogeneous Neumann condition (null flux for $\varphi$ at the boundary) \cite{chiarelli2017comparison}.

\subsection{Identification of parameters in the damage model}

The work presented in  \cite{boldrini2016non} addressed some mathematical deficiencies that have not been considered in previous descriptions.  Their model applies the phase field methodology to avoid limitations when dealing with crack initiation or branching and is based on the use of the basic principles of continuum mechanics.  Although this model has achieved good qualitative and quantitative results, one of its difficulties is the appearance of some materials parameters, with not necessarily a clear physical meaning, that needs to be found in order to reproduce a particular material behavior.

We are interested in the damage evolution in terms of parameters $\lambda_1$ $(g_c)$, $\lambda_2$ $(\lambda_c)$ and $\lambda_3$ $(\gamma_c)$, hence the damage evolution can be rewritten as:

\vspace{5mm}
\begin{equation}
\frac{\partial \varphi }{\partial t }=\Biggl( \frac{\lambda_1 \lambda_3}{\lambda_2}\Biggr)\frac{\partial }{\partial x }\Biggl( \frac{\partial\varphi  }{\partial x } \Biggr)+\Biggl( \frac{E}{\lambda_2}\Biggr) u_x^{2} (1-\varphi)-\Biggl( \frac{\lambda_1}{\lambda_2 \lambda_3 }\Biggr)\varphi.
\label{Eq10}
\vspace{5mm}
\end{equation}

We use pseudo-experimental data to train and validate the estimation procedure that will be proposed.  The pseudo-experimental data provides the evolution of the state variables of the model, $\varphi$  and $u$ , from the solution of the governing equations with known parameters.  The damage model is described by the system of governing differential equations \cref {Eq8} and \cref {Eq9} of hyperbolic and parabolic types, respectively. In both equations, the finite element method replaces the differential operator in space by a system of ordinary differential equations \cite{Fayed2018}. The assembled system is dependent on time and initial conditions in terms of  $u$  , $\dot u$ and $\varphi$   at time $t = 0$ in the domain.  At future times, the variables are approximated using the $\alpha$-Method for the parabolic equation and the Newmark method for the hyperbolic equation. 

The governing equations are coupled and therefore a strategy to relate their evolution progressively is necessary. An iterative semi-implicit scheme is implemented to accomplish this requirement. The  damage $\varphi$ at time step $n$ is used to solve \cref{Eq8} for the displacement $u$ at time step $n+1$ ; after that, the displacement  $u$  is used to calculate the strain $\frac{\partial u}{\partial x } $ and then \cref{Eq9} is solved. If the maximum damage in the model is less than 1, then the algorithm  returns to the first step. The iterations start with the initial damage $\varphi_0$ (at time $t=0$) as the first input to solve the displacement equation. Algorithm 1 resumes this  procedure.

\begin{algorithm}

\begin{algorithmic}
\STATE $\varphi \gets \varphi_0 $
\STATE $t \leftarrow 0 $
\WHILE{$ max(\varphi) < 1$}
\STATE $u \leftarrow$ Solve \cref {Eq8} using the FEM and the Newmark method
\STATE $\varphi \leftarrow$ Derive $u$ and solve \cref {Eq9} using the FEM and the $\alpha$- method
\STATE $t \leftarrow t+\Delta t $
\ENDWHILE
\end{algorithmic}
\caption{Iterative scheme to solve the governing equations }
\end{algorithm}

\section{Neural networks as function approximators}

Neural networks are mathematical constructs that are inspired by the brain capacity of humans and animals to perform complex tasks without much effort. However, they only have some similarities with the actual functioning of the brain that resembles the behavior and structure that humans have developed. A network is organized in layers made up of interconnected processing units where each connection is weighted and receives a  transformation by an activation function. The learning process is developed through experiences that are presented as training examples and the strength of the connections consolidates the knowledge acquired by the neural network.  Although this method is conceptually simple, it is possible to approximate non-linear relationships and complex patterns found in diverse applications \cite{remco_delft,Chakraverty2017}.

This work solves an identification problem based on the application of a feedforward neural network as a function approximator. A feedforward architecture is particularly adequate to approximate the physics model because it allows the explicit definition of space and time points in the domain as inputs of the neural network model. For many years the feedforward architecture has been explored in the solution of differential equations due to the possibility to compute analytic expressions of its derivatives using backward propagation \cite{Meade1994, Lagaris1998}. Recent advances in computational techniques, in particular automatic differentiation, have expanded this potential because now it is possible to define networks with multiple hidden layers, i.e., deep learning, and automatically compute their derivatives using the record of their operations \cite{raissi2017physics}. 

\subsection{Multilayer feedforward networks for function approximation tasks}\label{MFN}
 
Feedforward networks are typically fully connected, which means that every neuron in a layer is connected to every other neuron in the contiguous layer. Connections between the same or previous layers in the block are not allowed which means that there is no feedback communication during the forward computation of an output. The simplest feedforward structure is a single-layer network where the input layer directly feed their signals in the output. 

\begin{figure}[H]
\begin{center}
 
 \includegraphics[scale=0.7]{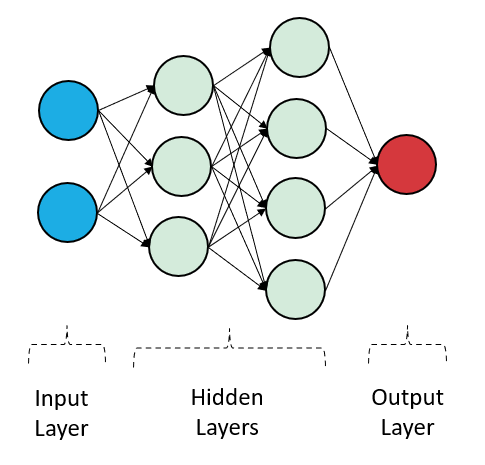} 
 \caption{\small {Architecture of a feedforward artificial neural network.}}
 \label{fig_ann}
\end{center}
\end{figure}

Multilayer feedforward networks add blocks of neurons between the input and output layers that are typically known as hidden layers. These layers help to detect relationships and patterns during the training. However, it is necessary to balance their number against the training time of the network. \cref{fig_ann} illustrates a feedforward neural network referred to as 2-3-4-1 because it has 2 inputs, 3 neurons in the first hidden layer, 4 neurons in the second hidden layer and finally 1 output. 

The basic structure of the neural networks selected for our application consists of an input layer that communicates with a block of one or more hidden layers using a system of weighted connections and biases. At the end of the network, the last hidden layer links to an output layer that provides an approximated value of the expected response. 
There are two simple operations applied to each neuron. The first operation is a weighted sum  for all the incoming values and an addition of a bias. Following that, an activation function $\sigma$ applies a nonlinear transformation which gives the actual output of the neuron. 
The strength of connections among neurons is defined by parameters $\theta$ (weights and biases) that are learned after a training process using labeled data. 

\begin{figure}[H]
\begin{center}
 
 \includegraphics[scale=0.7]{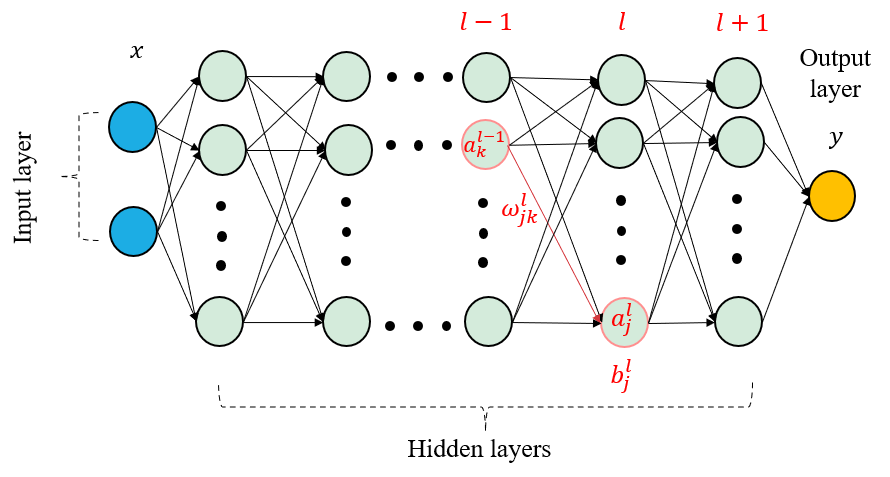} 
 \caption{\small {Multilayer feedforward architecture used in this work.}}
 \label{fig_nn}
\end{center}
\end{figure}

As presented in  \cref{fig_nn}, neurons in a given layer $l$ receive input from all the neurons in the previous layer $l-1$ and feed their output to all the neurons in the next layer $l+1$. 

The computation of the neural network output is known as forward propagation. The activation $a_j^{l}$ of the $j^{th}$ neuron of a layer $l$  is related to the activations in the layer $l-1$ by 
\vspace{5mm}
\begin{equation}
\label{eq11}
a_j^{l} = \sigma_l(\sum_{k}^{n^{l-1}}\omega_{jk}^{l}a_{k}^{l-1} +b_{j}^{l}),
\end{equation}

\noindent where the sum is over all neurons $k$ of the layer $l-1$ and the terms in this equation follow the notation used in  \cite{nielsen2015}: 

\begin{itemize}
   \item   $\omega_{jk}^{l}$ is  the weight for the connection from the $k_{th}$ neuron in the $(l-1)_{th}$ layer to the $j_{th}$ neuron of the  $l_{th}$ layer;
   \item  $b_{j}^{l}$ denotes the bias of the $j_{th}$  neuron of the $l_{th}$ layer;
   \item   $a_j^{l}$ represents the activation of the $j_{th}$  neuron in the $l_{th}$ layer;
   \item   $a_k^{l-1}$ represents the activation of the $k_{th}$  neuron in the $(l-1)_{th}$ layer.
   
\end{itemize}

\noindent A different activation function can be used in each layer and there are several options to apply this non-linear transformation. 

There is another important computation where the error signal passes leftward through the network. This left or backward pass is commonly known as backward propagation. It is a recursive process where the weights and bias of the neural network change in accordance with the sensitivity of the output to these parameters. 

\subsection{Approximate solution of differential equations using neural networks}\label{ASD}

The general mathematical form presented in \cref{Eq1} is now employed to represent the behavior of a dynamic system. After putting in evidence the variation with time and writing the dependencies of the operator $\mathcal{F}$ in a compact form, we have:
\begin{equation}
\frac{\partial u}{\partial t }= \mathcal{F}(u ,  \lambda ),\\
\label{Eq12}
\end{equation} \\
subject to the Dirichlet and Neumann boundary conditions
\vspace{5mm}
\begin{equation}
u(x,t)= g(x,t) , \qquad x \in \partial\Omega_D, 
\label{Eq13}
\end{equation} 
\begin{equation}
\frac{\partial u}{\partial x }=q(x,t), \qquad x \in \partial\Omega_N,
\label{Eq14}
\end{equation} \\
and with initial condition
\begin{equation}
t=0: \qquad u(x,0)= u_0(x) , \qquad x \in \Omega.  
\vspace{5mm}
\label{Eq15}
\end{equation}
Here $u(x,t)$ denotes the solution of the PDE,  $\mathcal F$ is a function or differential operator parametrized by $\lambda$ and $\Omega$ is the computational domain.

It is also important to define the residue of the PDE, which plays an important role in the training process. If we substitute an approximate solution $\tilde u$  in \cref{Eq12}  an error or residual $R$ exists,

\begin{equation}
R:= \frac{\partial \tilde u}{\partial t }- \mathcal F \bigl( \tilde u,\lambda \bigr). 
\label{Eq16}
\end{equation}

Following one of the first approaches proposed in \cite{Lagaris1998}, the approximated solution of a PDE can be constructed with a trial solution that satisfies the initial and boundary conditions. Consider the following trial solution:
\vspace{3mm}
\begin{equation}
u^{tr} =   G(x,t,NN(x,t, \theta)),
\label{Eq17}
\vspace{3mm}
\end{equation} 
\noindent where the term $NN(x,t,\theta)$ represents the output of a neural network model that takes as inputs points $(x,t)$ in the domain  $\Omega$ and its parametrized by learnable parameters $\theta$. The function $G(x,t)$ is defined to satisfy the boundary and initial conditions by construction.

In order to enforce the differential equation, the loss function $L$ of the neural network is defined in terms of the residue of the PDE and computed in a set of random points $(x_i,t_i)$ inside the domain as follows:
\vspace{5mm}
\begin{equation}
R_i (u^{tr})= \frac{\partial u^{tr}(x_i,t_i,\theta)}{\partial t }- \mathcal F \bigl(u^{tr},\lambda \bigr). 
\label{Eq18}
\end{equation} 
\begin{equation}
L =  \frac{1}{N_c} \sum_{i=1}^{N_c} f(0, R_i),
\label{eq19}
\vspace{5mm}
\end{equation} 
\noindent where $f$ represents a metric to measure the differences between the residue, i.e the target,  and their label or true value, i.e. zero, and $N_c$ is the number of collocation points. If we take the absolute value as the metric $f$, the expression given in \cref{eq19} can be interpreted as the collocation method of weighted residuals \cite{Fayed2018}. Nowadays, the terms in \cref{Eq18} are automatically computed in machine learning frameworks, but without the help of these tools it was necessary to derive the expressions to evaluate the residual function.

The physics informed methodology takes the trial solution  directly from the output of the neural network model,
\vspace{5mm}
\begin{equation}
u^{tr} = NN(x,t, \theta),
\label{Eq20}
\vspace{5mm}
\end{equation} 
\noindent and constructs the loss $L$ of the model  considering terms for the residue $L_r$, boundary conditions $L_b$ and initial conditions $L_i$,
\vspace{5mm}
\begin{equation}
L =  L_r + L_b + L_i,
\label{eq21}
\vspace{5mm}
\end{equation} 
\noindent where
\vspace{5mm}
\begin{equation}
L_r = \frac{1}{N_c} \sum_{i=1}^{N_c} f(0,R_i),
\label{eq22}
\vspace{5mm}
\end{equation} 
\begin{equation}
L_b = \frac{1}{N_b} \sum_{i=1}^{N_b} f(g_i,u_i^{tr}) + \frac{1}{N_b} \sum_{i=1}^{N_b} f \biggl(q_i,\frac{\partial u_i^{tr}}{\partial x }\biggr), 
\label{eq23}
\end{equation} 
\begin{equation}
L_i = \frac{1}{N_i} \sum_{i=1}^{N_i} f({u_0}_i,u_i^{tr}),  
\label{eq24}
\vspace{5mm}
\end{equation} 
\noindent $N_b$ and $N_i$  are the boundary and the initial conditions points, respectively. The label (true) values in equations \cref{eq23,eq24} are computed using the \cref{Eq13,Eq14,Eq15}.

In summary, a differential equation can be approximated employing directly the output of the neural network model as trial solution and enforcing the differential equation, boundary and initial conditions through the terms presented in \cref{eq21,eq22,eq23,eq24}. The training data is selected randomly inside the domain for $L_r$ and the samples for the other terms in \cref{eq21} are taken accordingly to the definition of the boundary and initial conditions for the dynamic system. During the training process, the learnable parameters $\theta$ are updated to minimize the total loss function until an optimization algorithm reach a maximum number of iterations or a convergence tolerance.

\subsection{Identification of parameters in differential equations using physics informed neural networks}\label{ADN}

The PINN methodology can also be employed in the discovery of the parameters $\lambda$ in the differential operator $ \mathcal F \bigl(u,\lambda \bigr)$ of \cref{Eq12}. Using some observations $\hat{u}_i$ (distributed in the space-time domain) and the partial differential equation (from a physics model), it is possible to estimate the parameters $\lambda$ that reproduces the physical behavior given by the data.

The physics informed model uses the same trial solution given in \cref{Eq20}, but now the residue of the PDE is a function of the output of the neural network and also of the unknown parameters $\lambda$ as
\vspace{5mm}
\begin{equation}
R_i(u^{tr} , \lambda)= \frac{\partial u^{tr}(x_i,t_i,\theta)}{\partial t }- \mathcal F \bigl(u^{tr},\lambda \bigr). 
\label{Eq25}
\vspace{5mm}
\end{equation} 
The parameters $\lambda$ that define the PDE equation of the model are learnable parameters similar to $\theta$, but they are external to the neural network. The loss function of the physics informed model is given by two contributions,
\vspace{5mm}
\begin{equation}
L =  L_r + L_c,
\label{Eq26}
\vspace{5mm}
\end{equation} 
\noindent where $L_r$ is the residue loss and $L_c$ is the collocation loss. In identification problems, the solution of the governing equations is partially known and the collocation loss can be interpreted as the criterion to fit the output of the neural network model to the pseudo-experimental data. 
Its training points $N_c$ are usually adopted as the same collocation points used for the residue loss and this loss is evaluated as follows:
\vspace{5mm}
\begin{equation}
L_c = \frac{1}{N_c} \sum_{i=1}^{N_c} f(\hat{u_i},u_i^{tr}). \label{eq27}
\vspace{5mm}
\end{equation} 
An important aspect of  this formulation is that the collocation loss  only affects the learning process of the parameters $\theta$. Conversely, the residue loss has influence in the adjustment of parameters $\theta$ and $\lambda$. 

\section{Methodology and Results}

The governing equations presented in \cref{damage model}  give the evolution of displacement and damage for a linear elastic material. In order to explore the use of neural networks in this model identification problem, different physical cases are considered following the assumptions taken for the final equations of the model.
The identification is primarily based on the damage evolution given that all the parameters from the phase field methodology appear in this expression. This consideration also reduces the complexity of the neural network model to only one parabolic partial differential equation.

We define our data-driven methodology using a feedforward neural network that takes space and time values as inputs and returns a continuous function of damage in the space-time domain. This output is employed to compute the residues of the partial differential equation and the boundary and initial conditions. 

As we presented in \cref{pid_sec}, there are different approaches to estimate parameters in differential equations. In most of these approaches,  authors have considered the fitting of the experimental data as the main objective of the problem and penalized the residue of the governing equations. We treat the identification of the parameters as a multi-objective optimization problem where it is necessary to achieve a good balance among each of the terms that appear in the loss function of our neural network model. In the physics informed methodology, the terms in the loss function are  simultaneously minimized without consideration of its role in the identification. This  can  increase the difficulty of material parameter estimation since the number of learning parameters in the neural network exceed by far the material parameters of the physics model. On the other hand, our implementation uses weights $\alpha_j$ for each term $L_j$ in the total loss expression $L$ stated as,
\vspace{5mm}
\begin{equation}
L = \alpha_r L_r +  \alpha_i L_i + \alpha_b L_b + \alpha_c L_c,
\vspace{5mm}
\label{eq28}
\end{equation}
\noindent where $L_r$, $L_i$, $L_b$ and $L_c$ are the residue, initial, boundary and collocation losses, respectively. It is natural to assume that the residue loss is more relevant in the estimation of the material parameters by its mathematical definition. While the residue loss is written from the governing equations of the model, the collocation loss is defined using the output of the neural network. This means that the corrections of the neural network parameters are directly affected by $L_c$ and, on the other hand, the material parameters depend heavily on $L_r$.

We use a few training simulations to tune the weights $\alpha_j$ considering their current contribution to the total loss. In most cases, this process is straightforward and produces good results. Other works based on the PINN classical model have already proposed the introduction of parameters to represent this relative importance of the loss function terms and observed significant improvements in their results \cite{remco_delft}. Another difference with the classical PINN methodology is that we include the boundary and initial conditions of the model because in some cases it facilitates the search of the neural network parameters. 

\paragraph{Hyperparameters and optimization strategy}
In neural network models, learnable parameters $\theta$ are progressively adjusted using the backpropagation algorithm. Hyperparameters, in contrast, are chosen using expert intuition. A typical search of hyperparameters includes the number of neurons and layers, loss and activation functions, amount of training data,  methods for weight initialization, batch size and learning rate of gradient descent, number of iterations kept in memory for the L-BFGS optimization, among others. 
The selection of hyperparameters defines the set of configurations that maximize a metric associated with the accuracy of a neural network model. We know from previous works that the order of the residue loss function is a metric closely related with the capacity of the network to approximate a PDE solution. However, in identification problems a low value of this metric does not guarantee a good estimation of the material constants $\lambda_j$.

After some tests using different configurations, we tune the hyperparameters using literature recommendations and practical experiences. We use the mean absolute function for the collocation loss $L_c$ and the mean square function for the residue $L_r$, boundary conditions $L_b$ and initial condition $L_i$ losses. We employ a tangent hyperbolic function as the default activation and the Glorot normal method to initialize the learnable weights of the model \cite{GlorotB10}. In contrast to common activation functions, the hyperbolic tangent is preferred because its non-linearity is complemented with a symmetrical distribution and greater derivatives. Furthermore, the Glorot initialization helps to avoid vanishing and exploding gradient problems.

For the optimization algorithms, we set the following hyperparameters:
\begin{itemize}
    \item Adam:  $\alpha = \SI{1e-3}{} , \beta_1=0.9 , \beta_2=0.999. $
    \item L-BFGS: 50 corrections for the limited memory and the convergence criterion to stop the iterations is $(L^k - L^{k+1})/max(|L^k|,|L^{k+1}|,1) \leq ftol$ , where $k$ is a given iteration and $ftol = \SI{1e-12}{}.$ 
\end{itemize}

During the tuning process, it was observed that the inclusion of the residue demands the use of networks with many layers and neurons to avoid poor estimates of the material constants. Possible reasons for this situation are local minima solutions, initial values defined for the search, bounds given in the optimization algorithms and hyperparameters used during the minimization. Although the penalizing constants regularize the search \cite{ramsay_parameter_2007}, we combine  some of the ideas of the two-step method \cite{Dua2011}, the principal differential analysis \cite{poyton_parameter_2006} and the generalized smoothing approach \cite{ramsay_parameter_2007} to address these difficulties. Our implementation employs the following three stages in the training:

\begin{itemize}

    \item \textbf{First stage}:  we use an initial L-BFGS algorithm with a maximum of 1000 iterations. In this stage, only the collocation loss $L_c$ is optimized and the aim is to refine the initial values of learnable parameters using the pseudo-experimental data available.

    \item \textbf{Second stage}: the second stage is a batch gradient descent optimization of $L$ using the Adam algorithm with a number of steps between 5000 and 20000. In this stage, we explore a combination of Adam and L-BFGS methods to reduce the occurrence of local minima solution. The strategy consists of training the network using the Adam algorithm, and after a defined number of iterations, performing the L-BFGS optimization of the collocation loss $L_c$ with a small limit of executions.

    \item \textbf{Third stage}: the training is complemented with a stage of simultaneous optimization using the L-BFGS method but this time with a maximum of 20000 iterations and a small tolerance for the convergence criterion. 

\end{itemize}

\noindent We observed that in some cases this strategy allowed the use of networks with a smaller number of layers and neurons, and consequently, this approach is applied in all cases. 

Some conjectures initially considered were that a higher amount of training data with refinement around the concentration of damage would provide better approximation results. However, it was noticed that an amount of approximately 10$\%$ of available data was enough to obtain good results without any improvement with larger percentages. Similarly, the results were not affected when the sampling had more points near the initial damage. 

Using the proposed optimization strategy in combination with the default configurations selected for most of the hyperparameters, it is possible to reduce the number of elements to be tuned. Finally, the number of neurons, layers, and the batch size of the Adam optimization are selected in a random search. This not only narrows the search but also lessens the necessity of expert intuition.

\paragraph{General methodology}
Though we use the core of the PINN methodology, we implemented some changes considering ideas from  works that apply neural networks to solve differential equations and also from estimation strategies that employ other function approximators. A short description of the principal elements of our implementation is summarized as follows:

\begin{enumerate}
 
  \item A deep feedforward neural network is created to compute the damage evolution $\varphi$. The network takes space $x$ and time $t$ as inputs.
  
  \item The output of this network is derived using automatic differentiation which allows the computation of the PDE residue $R$. The strain is part of the pseudo-experimental data generated using the forward solution of the partial differential equations.
  
  \item We use a loss expression $L$ composed of 4 important contributions:  collocation points, residue, boundary and initial conditions.
  
  \item  The contributions in the loss function are weighted using their relative importance $\alpha_j$. We interpret some terms as constraints of a multi-objective optimization process.
  
  \item The training process starts with a short optimization of the collocation loss $L_c$ and after this, a combination of Adam and L-BFGS is used to minimize the total loss expression $L$ and $L_c$ , respectively. Finally, we define an additional optimization stage where the neural network parameters $\theta$ are adjusted, and the material parameters of the governing equation $\lambda_j$ are identified simultaneously.
  
\end{enumerate}

\begin{figure}[H]
\centering
\includegraphics[width=0.8\linewidth]{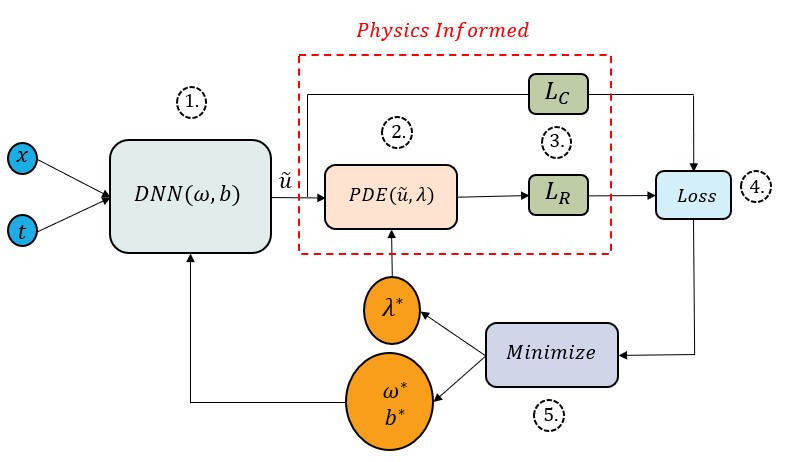}
\caption{PINN for parameter identification of the damage equation.}
\label{fig:pinn}
\end{figure}

The methodology  applied is represented in  \cref{fig:pinn}  with each part of the diagram related to the previous description. In spite of the fact that only the damage is approximated using the neural network, there is a coupling between damage and displacement equations that needs to be addressed. In the following subsections, we present how this relation was expressed and some results obtained.

\subsection{Identification  considering a constant strain in the bar}

The first implementation of PINNs for parameter identification is developed considering a decoupling of damage and displacement equations. In order to do that, it is assumed a constant strain in the spatial domain to compute the damage independently of the displacement evolution. This assumption reduces the mathematical complexity of the problem and gives the possibility to compare the solution with another identification approach.

\begin{figure}[H]
\centering
\includegraphics[width=0.5\linewidth]{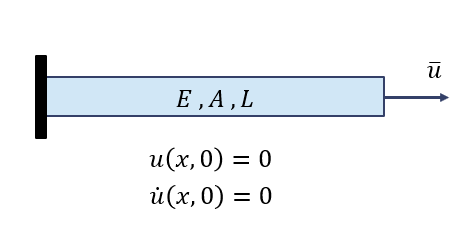}
\caption{Bar under constant normal strain.}
\label{fig:barra}
\end{figure}

The first case proposed is given by a bar fixed on the left side $(x=0)$  with zero initial displacement and velocity in the spatial domain which is shown in \cref{fig:barra}. A bar of length $L =$ \SI{1}{\meter}, cross sectional area $A =$ \SI{1e-3}{\metre\squared} and Young's modulus $E =$ \SI{72}{\giga\pascal} subjected to a constant strain of 1 $\%$ is considered.

\begin{figure}[H]
\centering
\includegraphics[width=0.5\linewidth]{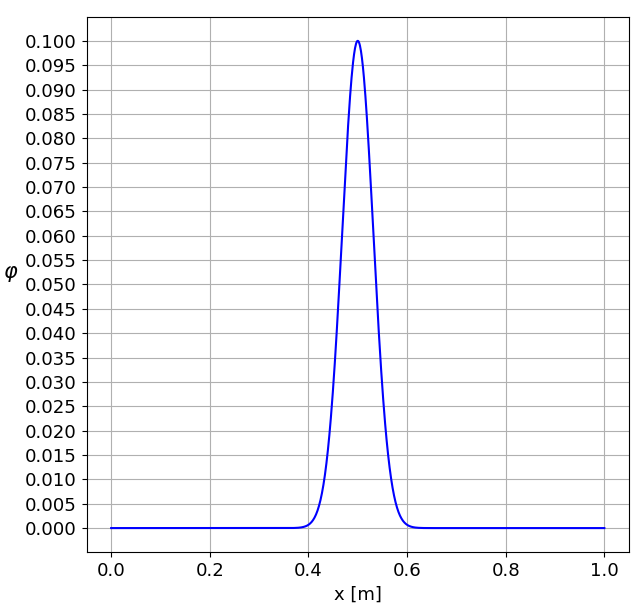}
\caption{Initial damage for the bar with the constant strain.}
\label{fig:initialdamage4.4}
\end{figure}

In the case of the damage equation, the boundary conditions are homogeneous Neuman conditions, and the initial condition presented in \cref{fig:initialdamage4.4} is a concentrated damage $\varphi_0=0.1$  in the middle of the bar $\Bigl(x= 0.5L\Bigr)$ introduced using a Gaussian function. 

The damage evolution in terms of parameters $\lambda_1$ $(g_c)$,  $\lambda_2$ $(\lambda_c)$ and $\lambda_3$ $(\gamma_c)$ can be rewritten as
\vspace{5mm}
\begin{equation}
\frac{\partial \varphi }{\partial t }=\Biggl( \frac{\lambda_1 \lambda_3}{\lambda_2}\Biggr)\frac{\partial }{\partial x }\Biggl( \frac{\partial\varphi  }{\partial x } \Biggr)+\Biggl( \frac{E^{*}}{\lambda_2}\Biggr) (1-\varphi)-\Biggl( \frac{\lambda_1}{\lambda_2 \lambda_3 }\Biggr)\varphi,
\label{Eq29}
\vspace{5mm}
\end{equation}
\noindent where $E^{*}$ represents the product of Young's modulus $E$ and the squared strain $u_x^{2}$.

The pseudo-experimental data for this identification is obtained from a forward solution using the methods and parameters given in \cref{table:1}.

\vspace{7pt}
\begin{table}[H]
\centering
\caption{Classical numerical solution for the bar with the constant strain.}
\label{table:1}
\vspace{6pt}
\begin{tabular}{cccccc}
\hline
Space discretization & Interpolation order  &  Nodes  & Time marching & $\Delta t$ & $t_f$\\\hline
FEM  & 1  & 1000  & Crank-Nicolson & \SI{2e-4}{\s} & \SI {0.303}{\s}
\\ \hline
\end{tabular}
\end{table}

The neural network was implemented using the open-source framework TensorFlow \cite{tensorflow2015-whitepaper} with the Python application software interface (API). The methodology implemented for material identification is based on a simple mathematical and computational structure that takes advantage of the tools and comprehensive libraries available in TensorFlow. 

\begin{table}[!ht]
\centering
\caption{Hyperparameters of the PINN for the bar under constant strain.}
\label{table:2}
\vspace{6pt}
\begin{tabular}{ccccccc}
\hline
Layers and neurons& Samples & L-BFGS  & Batch Size & Adam + L-BFGS &  L-BFGS  \\ \hline
[2,13,13,13,1]   &  5000 & 1000 & 2000 & 10000-80(500) & 20000 \\ \hline
\end{tabular}\end{table}

 \cref{table:2} presents the configurations that provided the best results. The architecture of the neural network is given by 2 inputs, 3 hidden layers with  13  neurons in each layer and an output layer with 1 neuron. Each stage of the optimization process has a defined number of maximum iterations and a total of  5000 $(x,t)$ pairs are used to compute the residue and collocation losses. The penalizing weights in the total loss function $L$ are $\alpha_r = 20$ , $\alpha_i = 10$, $\alpha_b = 2$, $\alpha_c= 1$. The first optimization stage employs the L-BFGS method, with a limit execution of 1000 iterations, to minimize the collocation loss. Then, in the second stage, a loop of 10000 iterations trains the total loss with the Adam method. Within this loop, each 500 iterations the collocation loss is minimized with an L-BFGS method with 80 executions at most. In the last stage the total loss is trained again with the L-BFGS method. The iteration limits of the first and second optimization stages are chosen empirically. These limits are lower considering that when the method alternates between the total and collocation losses there will be repetitive variations with a general decreasing trend. Thus, the role of these initial stages is to provide better estimates for the final optimization stage due to their trade-off in the search of the nuisance and structural parameters.

\begin{figure}[H]
\centering
\includegraphics[scale = 0.8]{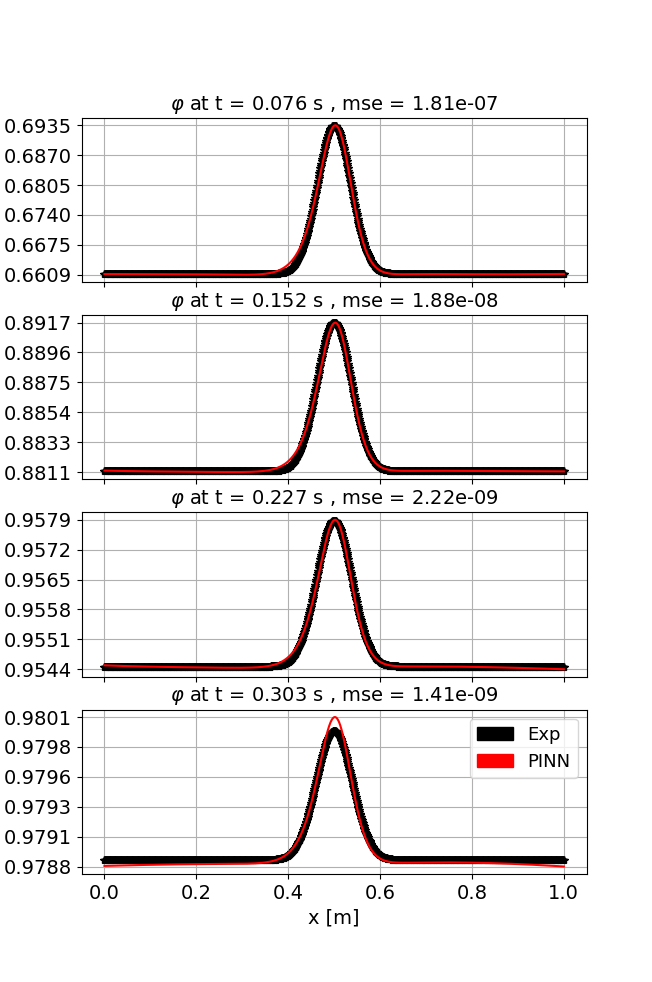}
\caption{Damage approximation for a bar under a constant strain.}
\label{fig:5}
\end{figure}

The percentage errors for the identification of the material parameters are presented in \cref{table:3} and the neural network approximation of damage at different times using this material constants is presented in \cref{fig:5}. The training time is about 8 minutes and the implementation ran in a GTX 1050 GPU.

\begin{table}[H]
\centering
\caption{Percentage errors in the identification using the PINN for the bar under constant strain.}
\label{table:3}
\vspace{6pt}
\begin{tabular}{cllc}
\hline
Parameter & Label value & Estimated value & Percentage error \\ \hline
$\lambda_1$  &  $\num{3.90e3}$  & $\hphantom{xy}\num{3.98e3}$  & $2.08 \%$ \\ 
$\lambda_2$  &  $\num{5.00e5}$  & $\hphantom{xy} \num{5.01e5}$ & $0.22 \%$ \\ 
$\lambda_3$  &  $\num{6.00e-2}$ & $\hphantom{xy}\num{6.11e-2}$  & $1.87 \%$ \\ \hline
\end{tabular}\end{table}

The results considering only the evolution of damage showed small percentage errors for all the parameters in the damage equation. Similarly, it can be seen in  \cref{fig:5} that the neural network approximation was accurate in comparison with the pseudo-experimental data provided. There were only slight differences in the last step ($t= 0.303$) of damage evolution but, in general, the neural network reproduced the qualitative and quantitative behavior given by the damage equation of the model.

Using this methodology, we considered a smaller number of hyperparameters in the tuning process, and with a total of 417 learnable parameters, it was possible to approximate the damage behavior and obtain a good estimation of the material parameters. 

\begin{figure}[H]
\centering
\includegraphics[scale = 0.6]{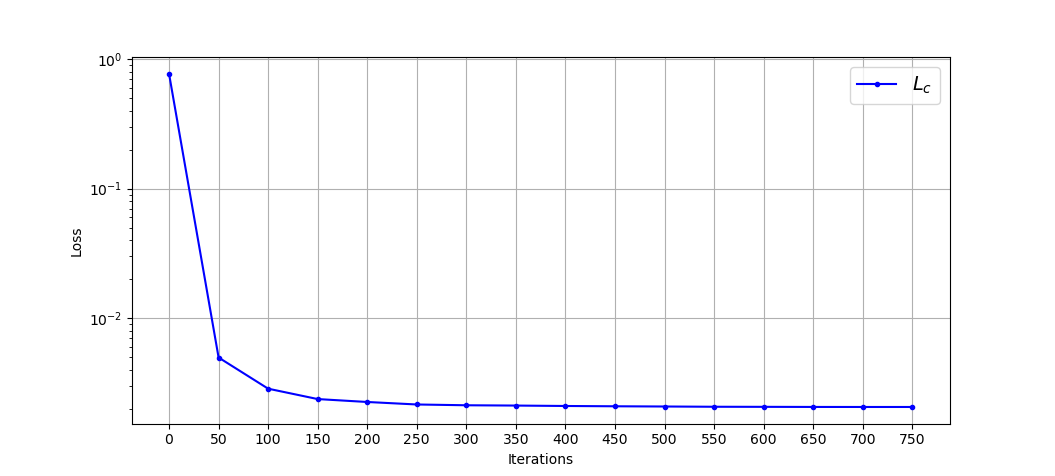}
\caption{Learning curve for the first optimization stage.}
\label{ops1}
\end{figure}

\begin{figure}[H]
\centering
\includegraphics[scale = 0.6]{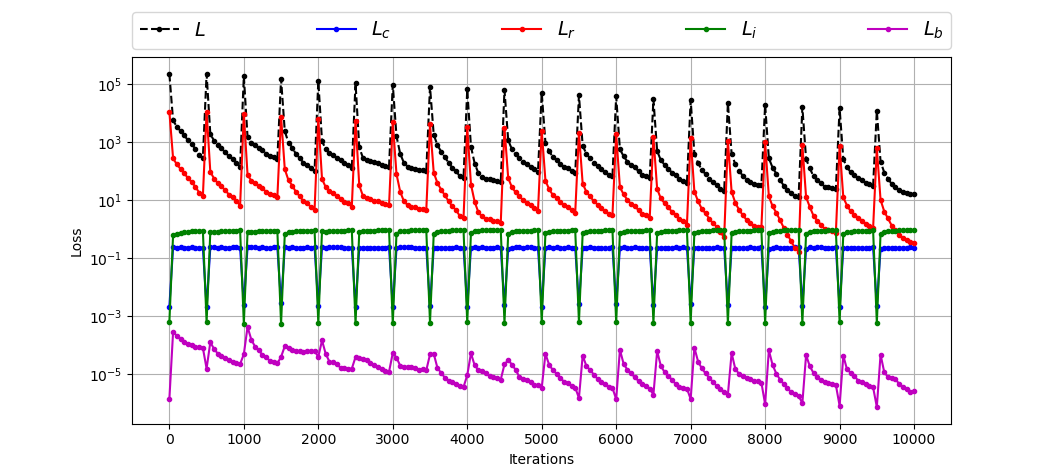}
\caption{Learning curve for the second optimization stage.}
\label{ops2}
\end{figure}

\begin{figure}[H]
\centering
\includegraphics[scale = 0.6]{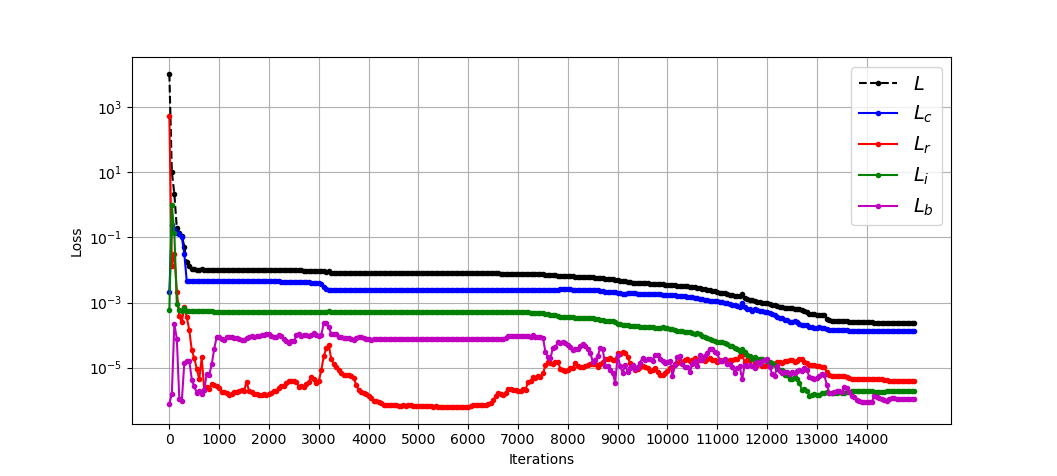}
\caption{Learning curve for the third optimization stage.}
\label{ops3}
\end{figure}

\cref{ops1,ops2,ops3}  present the learning curves for each optimization stage. As described previously, the metric for the collocation loss is the mean absolute error and the mean square function is used for the other terms.

\cref{ops1} shows how the collocation loss plunges in the initial 50 iterations of the first stage and then remains steady. The seasonality observed in \cref{ops2} is caused by the execution of the L-BFGS method (every 500 iterations) to minimize only the collocation loss. It can be noticed that the peaks in the total loss $L$ coincide with the troughs of the collocation loss $L_c$. Although at the end of the second stage the total loss is still high, there is a decreasing trend and a considerable reduction during this stage.

\cref{ops3} presents the last training stage of the proposed methodology. All the curves fall in the initial iterations, but after this drop, most of them remain fairly unchanged in a long region. The curves decline at the end of the optimization followed by a short plateau, where the method reaches the stopping criterion. The total loss is driven mainly by the behavior of the collocation loss and the boundary, initial and residue losses end around the same values.

\subsubsection{PDE constrained optimization in FEniCS }

As part of verification and validation, an identification in the FEniCS system was implemented using the same considerations previously stated. This framework has an intuitive mathematical interface, express problem solution using a high-level syntax and maybe its most important feature is that its code generation technology generates parallel optimized low-level C++ code for the solution of forward and adjoint systems. As a consequence of this optimization in the code, it is possible to compute functional and gradient information easily and pass this information to the optimization algorithm.

The computation of the gradient using functional perturbations, finite differences, and other approximation methods are affected by the propagation of errors or expensive computations. Alternatively, the adjoint method computes the gradient of a scalar function with a cheaper procedure similarly to automatic differentiation and requires only one PDE evaluation. As presented in \cite{funke2013framework}, the user describes the forward model, the control parameters and the objective function using a high-level syntax called UFL. The optimization framework then repeatedly re-executes the tape (record of operations used in the solution of the equations ) to evaluate the functional value, solves the adjoint PDE to compute the functional gradient, and modifies the tape to update the control parameters until an optimal solution is found. Details about how the adjoint equation is derived using the first-order optimality conditions for a PDE constrained optimization problem can be found in \cite{funke2013framework}. 

The identification with FEniCS adjoint employs the damage evolution in terms of parameters $\lambda_1^* ,  \lambda_2^* $ and $\lambda_3^*$  as

\vspace{5mm}
\begin{equation}
\frac{\partial \varphi }{\partial t }=\lambda_1^*\frac{\partial }{\partial x }\Bigl( \frac{\partial\varphi  }{\partial x } \Bigr)+\lambda_2^* (1-\varphi)-\lambda_3^*\varphi,
\label{eq30}
\vspace{5mm}
\end{equation}
where,
\vspace{5mm}
\begin{equation}
\lambda_1^* = \frac{\lambda_1\lambda_3}{\lambda_2 }
\label{eq31}
\end{equation}
\begin{equation}
\lambda_2^* = \frac{E^*}{\lambda_2 }
\label{eq32}
\end{equation}
\begin{equation}
\lambda_3^* = \frac{\lambda_1}{\lambda_3\lambda_2},
\vspace{5mm}
\label{eq33}
\end{equation}
and the results of the estimation are presented in \cref{table:4} using $\lambda_1 ,\lambda_2,\lambda_3$ for comparison. The running time was about 15 minutes in a computer with processor Intel(R) Core(TM) i5-8300H CPU with memory of 12GB RAM.

\vspace{7pt}
\begin{table}[!ht]
\centering
\caption{Percentage error in the identification using FEniCS for a bar with constant strain.}
\label{table:4}
\vspace{6pt}
\begin{tabular}{cllc}
\hline
Parameter & Label value & Estimated value & Percentage error \\ \hline
$\lambda_1$  &  $\num{3.90e3}$ & $\hphantom{xy} \num{3.89e3}$  & $0.025 \%$ \\ 
$\lambda_2$  &  $\num{5.00e5}$ & $\hphantom{xy} \num{4.99e5}$ & $0.020 \%$ \\ 
$\lambda_3$  &  $\num{6.00e-2}$ & $ \hphantom{xy} \num{5.99e-2}$  & $0.017 \%$ \\
\end{tabular}\end{table}

In this case, the pseudo-experimental data was obtained using a mesh with 1000 nodes and the functional was constructed with samples in the time domain and considering all the spatial values for a given sample.
5 time steps of a total of 1527 were randomly selected and used to construct the objective function that was minimized. 

The identification using this framework gave significant smaller errors for the material parameters in comparison with the PINN methodology. However, the automated PDE constrained optimization required all the spatial information for each time step sample used in the estimation and was performed in terms of linear parameters $\lambda_1^* ,\lambda_2^*,\lambda_3^*$ . In this respect, the neural network approach is more general and can be easily adapted to experimental settings where all the spatial information is not available. Another important remark is that the use of function approximators has proven to be better to estimate non-linear parameters in differential equation models and also has good performance in the presence of noise. On the other hand, in solutions with classical numerical methods, such as the finite element method, the number of material parameters increases with the refining of the mesh in non-linear problems and some numerical issues can be developed in the presence of noisy data \cite{berg2017neural}.

\subsubsection{ Noise robustness of the methods}

One of the advantages of neural networks as function approximators is that they have shown to be robust in the presence of noise \cite{raissi2017physics, borodinov_deep_2019}. To asses this, we perform the identification of the parameters using 20 different values of uncorrelated noise between 0 and $10 \%$ similar to the systematic study presented in \cite{raissi2017physics}. The parameters estimated for different levels of noise are in \cref{table:5}  and \cref{table:6} presents the mean, the standard deviation and the relative standard deviation of the identification to evaluate the robustness of the implementation to the noise.

\vspace{7pt}
\begin{table}[!ht]
\centering
\caption{Parameters estimated for different levels of noise using a PINN.}
\label{table:5}
\vspace{6pt}
\begin{tabular}{ccccc}
\hline
\text{Level of noise \%} & $\lambda_1$ &  $\lambda_2$  & $\lambda_3$\\ \hline
0.0   &  3772.955 &  501054.931 &     0.058 \\
0.5   &  3913.628 &  501067.868 &     0.060 \\
1.0   &  3757.971 &  503823.802 &     0.065 \\
1.5   &  3771.606 &  500928.572 &     0.058 \\
2.0   &  3663.080 &  503543.274 &     0.063 \\
2.5   &  3765.409 &  503434.252 &     0.065 \\
3.0   &  3861.293 &  500984.546 &     0.059 \\
3.5   &  3805.192 &  503100.513 &     0.064 \\
4.0   &  3856.346 &  502600.354 &     0.063 \\
4.5   &  4116.996 &  500612.807 &     0.061 \\
5.0   &  3855.899 &  502553.604 &     0.063 \\
5.5   &  3625.063 &  500640.227 &     0.054 \\
6.0   &  3891.068 &  502382.433 &     0.063 \\
6.5   &  3596.138 &  500395.233 &     0.053 \\
7.0   &  4186.175 &  500603.370 &     0.063 \\
7.5   &  3520.839 &  500572.652 &     0.053 \\
8.0   &  3913.673 &  502100.357 &     0.063 \\
8.5   &  3866.172 &  502254.003 &     0.063 \\
9.0   &  3945.199 &  501774.604 &     0.062 \\
9.5   &  3971.563 &  500254.033 &     0.058 \\
10.0  &  3453.295 &  500429.881 &     0.051 \\
\end{tabular}\end{table}

\vspace{7pt}
\begin{table}[!ht]
\centering
\caption{Influence of noise in the estimation of the parameters for a PINN.}
\label{table:6}
\vspace{6pt}
\begin{tabular}{cllcc}
\hline
Parameter &  Label value & Mean value  &  Relative standard deviation & Mean error\\ \hline
$\lambda_1$  &  $\hphantom{x} \num{3.90e3}$  & $\hphantom{x} \num{3.81e3}$  & $4.57 \%$  & $3.77 \%$\\ 
$\lambda_2$  &  $\hphantom{x} \num{5.00e5}$  & $\hphantom{x} \num{5.02e5}$ & $0.23 \%$  & $0.33 \%$\\ 
$\lambda_3$  &  $\hphantom{x} \num{6.00e-2}$ & $\hphantom{x}\num{6.01e-2}$  & $7.06 \%$   & $6.02 \%$ \\ \hline
\end{tabular}\end{table}

We let the same settings to all the levels of noise but we know that it could exist a better combination of hyperparameters for each of them. From the quantitative results in \cref{table:5} we see that the estimation of the parameter $\lambda_2$ is very accurate and the results for the other parameters are within an acceptable range.

We applied the same test of robustness for the solution using constrained optimization in FEniCS and the results are shown in \cref{table:7,table:8}. Despite having some good estimates for mean values, we observe that the relative standard deviation and the mean error are higher in contrast to the results of the neural network methodology.

\vspace{7pt}
\begin{table}[!ht]
\centering
\caption{Parameters estimated for different levels of noise in FEniCS}
\label{table:7}
\vspace{6pt}
\begin{tabular}{ccccc}
\hline
\text{Level of noise} & $\lambda_1$ &  $\lambda_2$  & $\lambda_3$\\ \hline
0.0 \%  &  3899.963 &  499999.990 &     0.060 \\
0.5 \%  &  3841.518 &  500037.703 &     0.059 \\
1.0 \%  &  3782.259 &  500075.429 &     0.058 \\
1.5 \%  &  3722.152 &  500113.166 &     0.057 \\
2.0 \%  &  3661.160 &  500150.915 &     0.056 \\
2.5 \%  &  3599.243 &  500188.677 &     0.055 \\
3.0 \%  &  3536.359 &  500226.450 &     0.055 \\
3.5 \%  &  3472.442 &  500264.313 &     0.054 \\
4.0 \%  &  3407.519 &  500302.095 &     0.053 \\
4.5 \%  &  3341.461 &  500339.888 &     0.052 \\
5.0 \%  &  3274.204 &  500377.696 &     0.051 \\
5.5 \%  &  3205.683 &  500415.521 &     0.050 \\
6.0 \%  &  3135.826 &  500453.362 &     0.048 \\
6.5 \%  &  3064.554 &  500491.217 &     0.047 \\
7.0 \%  &  2991.255 &  500529.097 &     0.046 \\
7.5 \%  &  2916.964 &  500566.998 &     0.045 \\
8.0 \%  &  2841.060 &  500604.904 &     0.044 \\
8.5 \%  &  2763.269 &  500642.795 &     0.043 \\
9.0 \%  &  2683.043 &  500680.681 &     0.042 \\
9.5 \%  &  2600.532 &  500718.606 &     0.040 \\
10.0 \% &  2515.914 &  500756.562 &     0.039 \\
\end{tabular}\end{table}

\vspace{7pt}
\begin{table}[!ht]
\centering
\caption{Influence of noise in the estimation of the parameters for constrained optimization.}
\label{table:8}
\vspace{6pt}
\begin{tabular}{cllcc}
\hline
Parameter &  Label value & Mean value  &  Relative standard deviation & Mean error\\ \hline
$\lambda_1$  &  $\hphantom{x} \num{3.90e3}$  & $\hphantom{x} \num{3.25e3}$  & $12.82 \%$  & $16.66 \%$\\ 
$\lambda_2$  &  $\hphantom{x} \num{5.00e5}$  & $\hphantom{x} \num{5.01e5}$ & $0.05 \%$  & $0.08 \%$\\ 
$\lambda_3$  &  $\hphantom{x} \num{6.00e-2}$ & $\hphantom{x} \num{5.02e-2}$  & $12.58 \%$   & $16.34 \%$ \\ \hline
\end{tabular}\end{table}

\subsection{ Identification considering the displacement evolution}

After exploring the use of PINNs without considering the displacement evolution, now we return to the coupled system of equations for the identification. In this case, the evolution of displacement is given as an input to the neural network making the inverse analysis more consistent with the real behavior. 

We propose a total of four cases with three distinct initial conditions for damage in each of them. The neural network identifies a different set of material parameters in problems with several boundary and initial conditions. With these cases, we evaluate the generalization capabilities and robustness of the implemented methodology.

In all cases, we adopt a bar of length  $L =$ \SI{1}{\meter} , cross sectional area  $A =$ \SI{1e-3}{\metre\squared}, density   $\rho =$ \SI{2810}{\kilo\gram\per\cubic\metre} and Young's modulus $E =$ \SI{72}{\giga\pascal}. The initial conditions for the displacement equation are  $u=0$ and $\dot{u}=0$ in the spatial domain.

The pseudo-experimental data used for the identification gives the evolution of damage in its domain and includes the strain distribution in the bar for the learning process. The classical methods used to obtain the training are summarized in \cref{table:9}.

\begin{table}[!ht]
\centering
\caption{Classical numerical solution for the cases with displacement evolution.}
\label{table:9}
\vspace{6pt}
\begin{tabular}{cccccc}
\hline
Space discretization & Interpolation order - Nodes &  & Time marching & $\Delta t$ \\ \hline
FEM  &2-1000 & & $\alpha$-Method / Newmark & \SI{1e-4}{\s} 
\\ \hline
\end{tabular}
\end{table}

\subsubsection{Case 1}

\begin{figure}[h]
\centering
\includegraphics[width=1\linewidth]{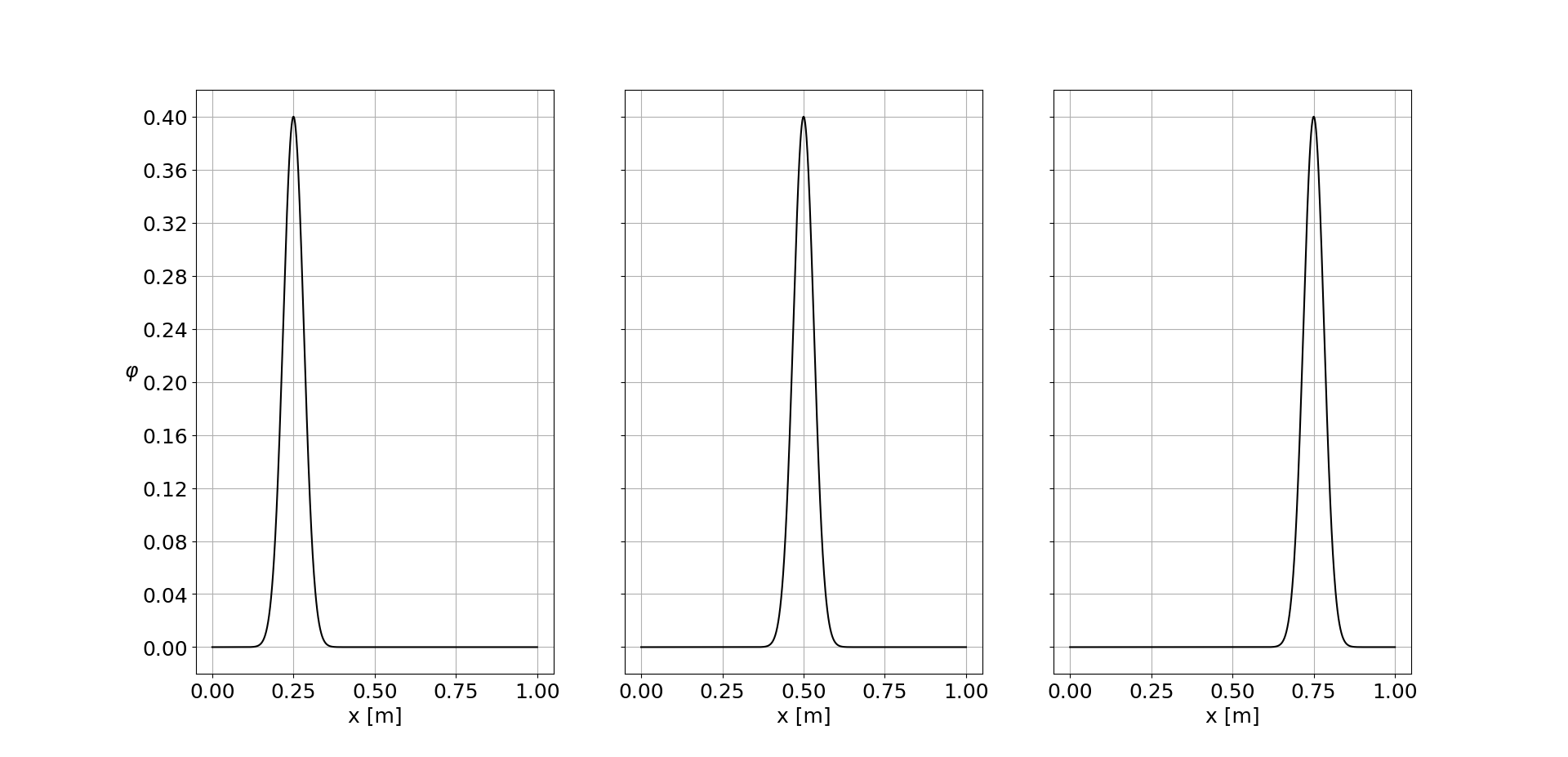}
\caption{Initial conditions considered for case 1 with $\varphi_0 = 0.4$.}
\label{fig:initial_damage}
\end{figure}

In this case, the bar is fixed at the end $x=0$ and subjected to a displacement $u_L = \SI{1e-3}{\meter}$  at the end $x=L$. The three initial conditions tested  for damage are presented in \cref{fig:initial_damage}. 

\noindent The damage evolution is rewritten in terms of  $\lambda_1 ,  \lambda_2 $ and $\lambda_3$ as
\vspace{5mm}
\begin{equation}
\frac{\partial \varphi }{\partial t }=\Biggl( \frac{\lambda_1 \lambda_3}{\lambda_2}\Biggr)\frac{\partial }{\partial x }\Biggl( \frac{\partial\varphi  }{\partial x } \Biggr)+\Biggl( \frac{E}{\lambda_2}\Biggr) (1-\varphi)\Biggl(\frac{\partial u}{\partial x } \Biggr)^{2}-\Biggl( \frac{\lambda_1}{\lambda_2 \lambda_3 }\Biggr)\varphi.
\label{Eq34}
\vspace{5mm}
\end{equation}
\noindent with the strain calculated from the solutions of the displacement equation
\vspace{5mm}
\begin{equation}
\rho \frac{\partial^2 u }{\partial t^2 }=\frac{\partial }{\partial x }\Bigl((1-\varphi)^{2} E\frac{\partial u}{\partial x }\Bigr).\label{Eq35}
\vspace{5mm}
\end{equation}
The use of both state variables $u$ and $\varphi$ required some minor changes in the depth and width of the neural network. The hyperparameters adopted for the neural network are presented in  \cref{table:10}, the penalizing weights used in $L$ are $\alpha_r = 10$ , $\alpha_i = 8$, $\alpha_b = 2$, $\alpha_c= 10$, and the same configuration is used for the three initial conditions. 

\begin{figure}[H]
\centering
\includegraphics[width= 0.99\linewidth]{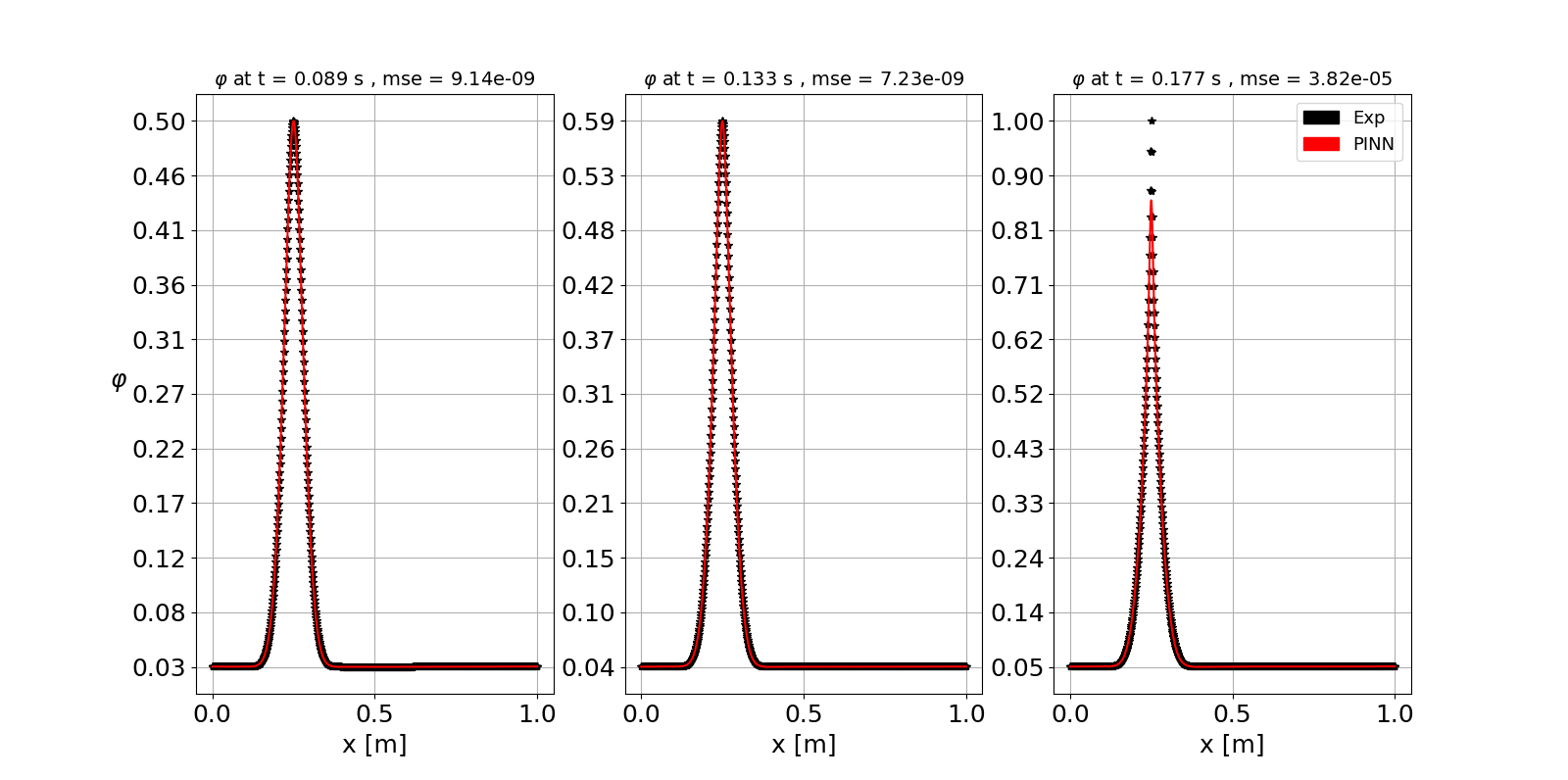}
\caption{Case 1 with initial condition $\varphi_0^{max}$ $ at $ $x_0=0.25.$}
\label{c1i1}
\end{figure}

\begin{figure}[H]
\centering
\includegraphics[width=0.99\linewidth]{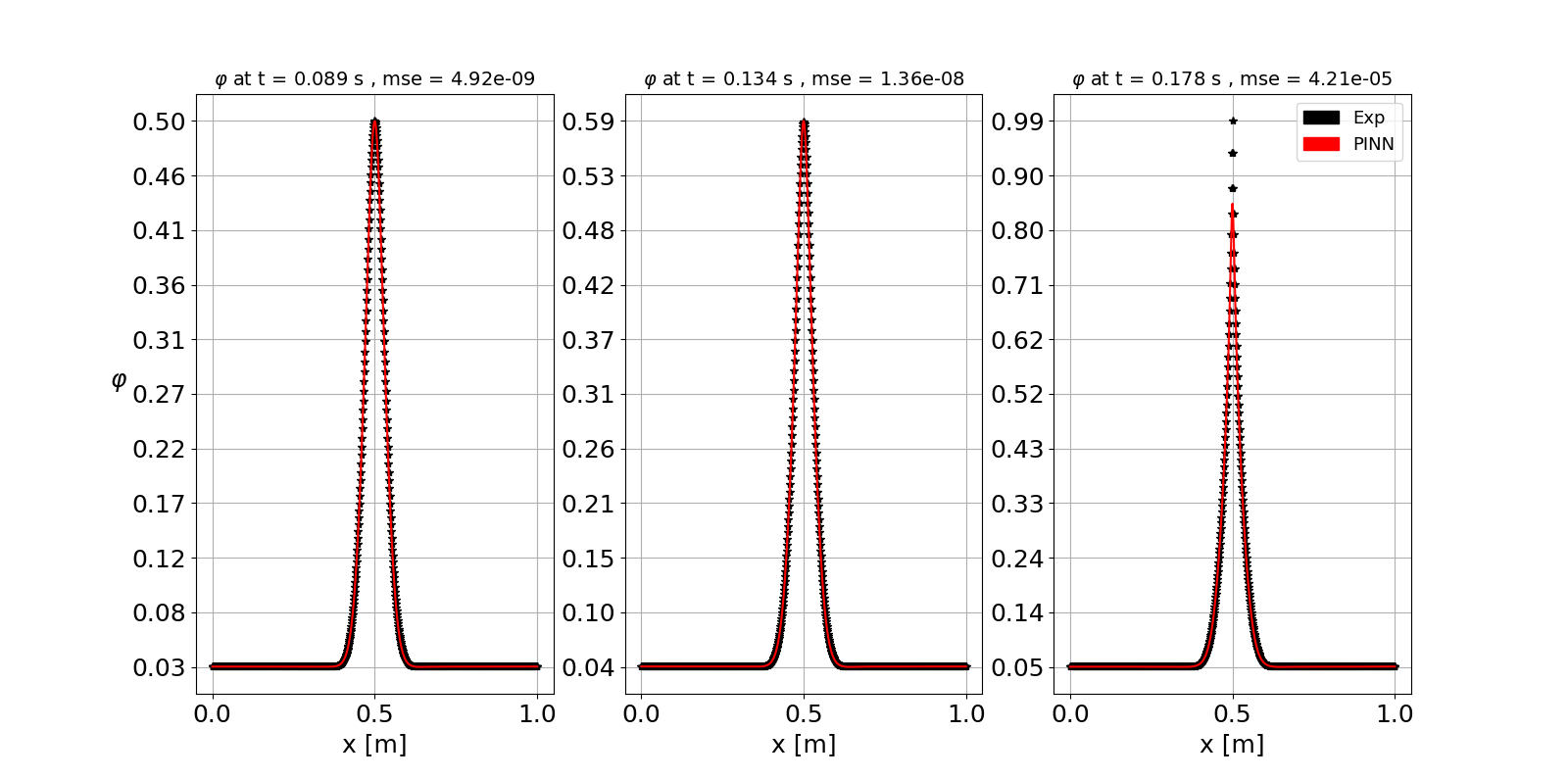}
\caption{Case 1 with initial condition $\varphi_0^{max}$ $ at $ $x_0=0.5.$}
\label{c1i2}
\end{figure}

\begin{figure}[H]
\centering
\includegraphics[width=0.99\linewidth]{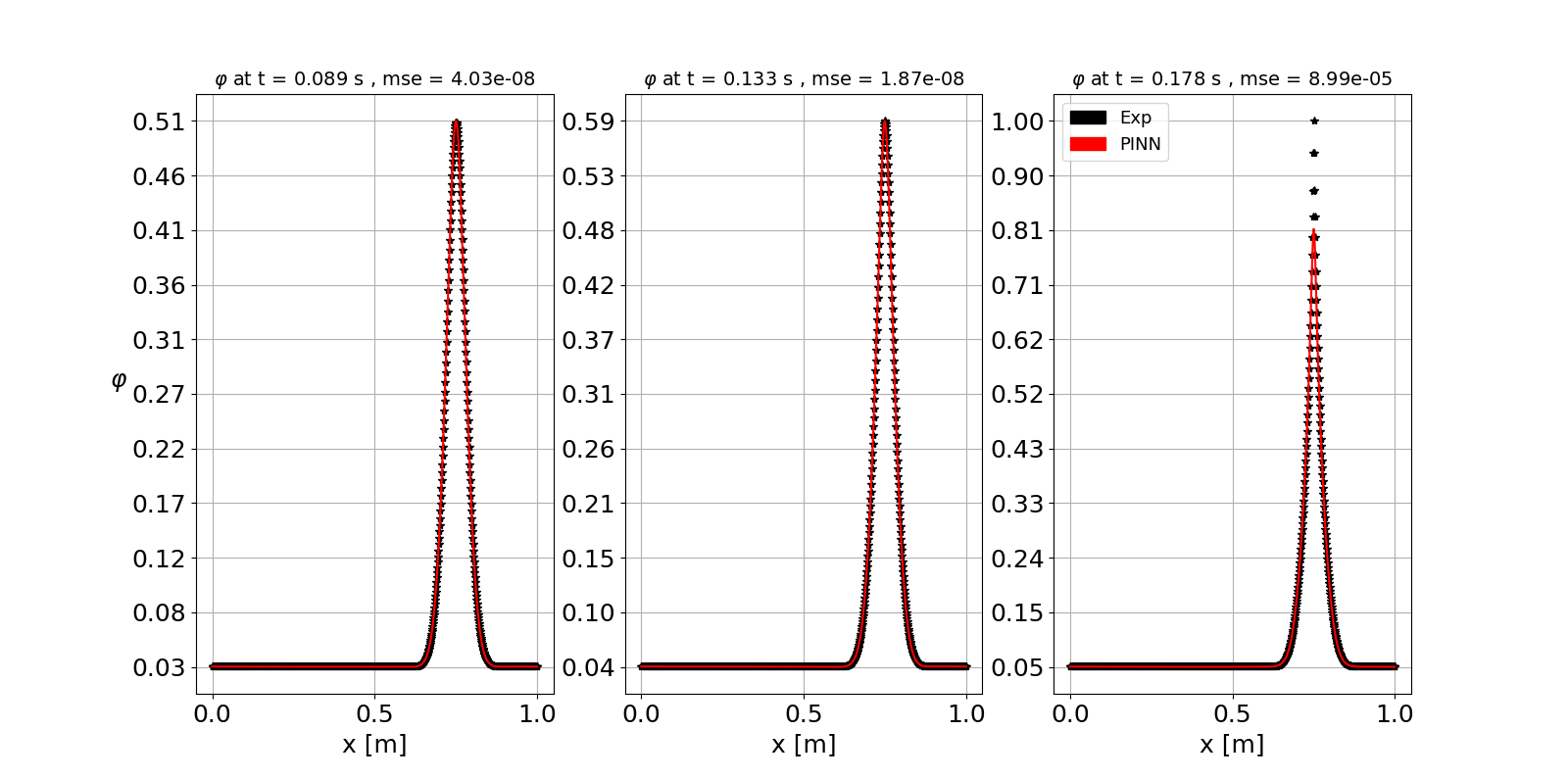}
\caption{Case 1 with initial condition $\varphi_0^{max}$ $ at $ $x_0=0.75.$}
\label{c1i3}
\end{figure}

\begin{table}[!ht]
\centering
\caption{Hyperparameters of the PINN for case 1.}
\label{table:10}
\vspace{6pt}
\begin{tabular}{ccccccc}
\hline
Layers and neurons& Samples & L-BFGS  & Batch Size & Adam + L-BFGS &  L-BFGS  \\ \hline
[2,25,25,25,25,1]   &  5000 & 1000 & 2000 & 10000-80(500) & 20000 \\ \hline
\end{tabular}\end{table}
The  estimation of the material parameters was performed using an exponential scale and defining bounds with enough range for the parameter search. The percentage errors for the parameters are given in \cref{table:11} and the PINN approximation is presented in \cref{c1i1,c1i2,,c1i3}.

The material parameters estimated using the physics informed neural network methodology provided good results with percentage errors below a $7\%$. In general, the errors in the identification were smaller for the parameter $\lambda_2$ and had similar values for the others. It is also important to remark that the same configuration was used for all the initial conditions but it is possible to optimize each of them if a smaller error is desired.

\begin{table}[H]
\centering
\caption{Percentage errors in the identification for case 1.}
\label{table:11}
\vspace{6pt}
\begin{tabular}{ccllc}
\hline
              &  Parameter & Label value & Estimated value & Percentage error\\ \hline
              
\multirow{3}{6em}{$\varphi_0^{max}$ $ at $ $x_0=0.25$} 
& $\lambda_1$
& $\num{4.00e3}  $ & $ \hphantom{xy} \num{3.76e3}$  & $5.95 \%$ \\ 
& $\lambda_2$  
& $\num{1.60e5}  $ & $ \hphantom{xy} \num{1.69e5}$  & $5.40 \%$ \\ 
& $\lambda_3$
& $\num{2.00e-2} $ & $ \hphantom{xy} \num{2.14e-2}$  & $6.93 \%$ \\ \hline

\multirow{3}{6em}{$\varphi_0^{max}$ $ at $ $x_0=0.5$}  
& $\lambda_1$
& $\num{4.00e3}  $ & $ \hphantom{xy} \num{3.87e3}$  & $3.25 \%$ \\ 
& $\lambda_2$
& $\num{1.60e5}  $ & $ \hphantom{xy} \num{1.64e5}$  & $2.41 \%$ \\ 
& $\lambda_3$
& $\num{2.00e-2} $ & $ \hphantom{xy} \num{2.06e-2}$  & $2.88\%$ \\ \hline

\multirow{3}{6em}{$\varphi_0^{max}$ $ at $ $x_0=0.75$}  
& $\lambda_1$
& $\num{4.00e3}$  & $ \hphantom{xy} \num{3.78e3}$  & $5.50 \%$ \\ 
& $\lambda_2$
& $\num{1.60e5}$  & $ \hphantom{xy} \num{1.67e5}$  & $4.26 \%$ \\ 
& $\lambda_3$
& $\num{2.00e-2}$ & $ \hphantom{xy} \num{2.09e-2}$  & $4.48 \%$ \\ \hline
\end{tabular}\end{table}


In \cref{c1i1,c1i2,c1i3}, we observe that for all the initial conditions proposed the damage has a similar behavior. 
The differences between the target values and the output of the neural network are measured using the mean square error (mse) and the PINN solution follows the expected evolution in the figures presented. However, for values close to the final time step of the experimental data the output of the neural network seems to be delayed. This can be a result of a smoothed solution of the neural networks in this closed region.

\subsubsection{Case 2}

In this case, the bar is fixed at the end $x=0$ and subjected to a force $F_L = \SI{100} {\kilo\newton}$  at the end $x=L$. The initial conditions tested  for the damage are applied using a
Gaussian distribution centered at $x = 0.25$ , $x = 0.5$ and $x = 0.75$ with  $\varphi_0 = 0.3$. Although this case is physically equivalent to the previous, it considers a Neuman boundary condition for the displacement equation and a different $\varphi_0$ for the initial damage.

The hyperparameters adopted for the neural network are presented in  \cref{table:12}. The first row has the configurations for the initial conditions centered at  $x = 0.25$ and  $x = 0.5$. It was necessary to increase the amount of neurons for the initial condition centered at  $x = 0.75$ as presented in the second row. For this and the following two cases, the penalizing weights used in the total loss function are $\alpha_r = 20$ , $\alpha_i = 8$, $\alpha_b = 2$, $\alpha_c= 15.$
\begin{table}[H]
\centering
\caption{Hyperparameters of the PINN for case 2.}
\label{table:12}
\begin{tabular}{ccccccc}
\hline
Layers and neurons& Samples & L-BFGS  & Batch Size & Adam + L-BFGS &  L-BFGS  \\ \hline
[2,14,14,14,14,14,1]   &  8000 & 1000 & 2000 & 10000-80(500) & 20000 \\\hline
[2,18,18,18,18,18,1]   &  8000 & 1000 & 2000 & 10000-80(500) & 20000 \\ \hline
\end{tabular}
\end{table}
The percentage errors of the parameter identification are presented in \cref{table:13}. For this physical case, we had a larger error in $\lambda_3$ and a smaller error in $\lambda_1$ for all the initial conditions proposed. This could be associated with the sensitivity of the damage response to these constants.  In contrast to the previous case, we employed more neurons for the last initial condition considered to have an acceptable percentage error for $\lambda_3$. 
\begin{table}[H]
\centering
\caption{Percentage error in the identification for case 2.}
\label{table:13}
\vspace{6pt}
\begin{tabular}{ccllc}
\hline
              &  Parameter & Label value & Estimated value & Percentage error\\ \hline
\multirow{3}{6em}{$\varphi_0^{max}$ $ at $ $x_0=0.25$}  
& $\lambda_1$               
&  $\num{8.00e3} $ & $ \hphantom{xy} \num{7.99e3}$  & $0.06 \%$ \\ 
& $\lambda_2$
&  $\num{2.00e5}  $ & $ \hphantom{xy} \num{2.09e5}$  & $4.50 \%$ \\ 
& $\lambda_3$
&  $\num{1.00e-2} $ & $ \hphantom{xy} \num{1.06e-2}$  & $6.40 \%$ \\ \hline

\multirow{3}{6em}{$\varphi_0^{max}$ $ at $ $x_0=0.5$}
& $\lambda_1$
&  $\num{8.00e3}  $ & $ \hphantom{xy} \num{7.94e3}$  & $0.72 \%$ \\ 
& $\lambda_2$
&  $\num{2.00e5}  $ & $ \hphantom{xy} \num{2.10e5}$  & $4.82 \%$ \\ 
& $\lambda_3$
&  $\num{1.00e-2} $ & $ \hphantom{xy} \num{1.06e-2}$  & $6.29\%$ \\ \hline

\multirow{3}{6em}{$\varphi_0^{max}$ $ at $ $x_0=0.75$}
& $\lambda_1$
&  $\num{8.00e3}$ & $ \hphantom{xy} \num{8.17e3}$  & $2.08 \%$ \\ 
& $\lambda_2$  
&  $\num{2.00e5} $ & $ \hphantom{xy} \num{2.10e5}$  & $5.18 \%$ \\ 
& $\lambda_3$
&  $\num{1.00e-2} $ & $ \hphantom{xy} \num{1.10e-2}$  & $9.66 \%$ \\ \hline

\end{tabular}\end{table}

\begin{figure}[H]
\centering
\includegraphics[width=0.9\linewidth]{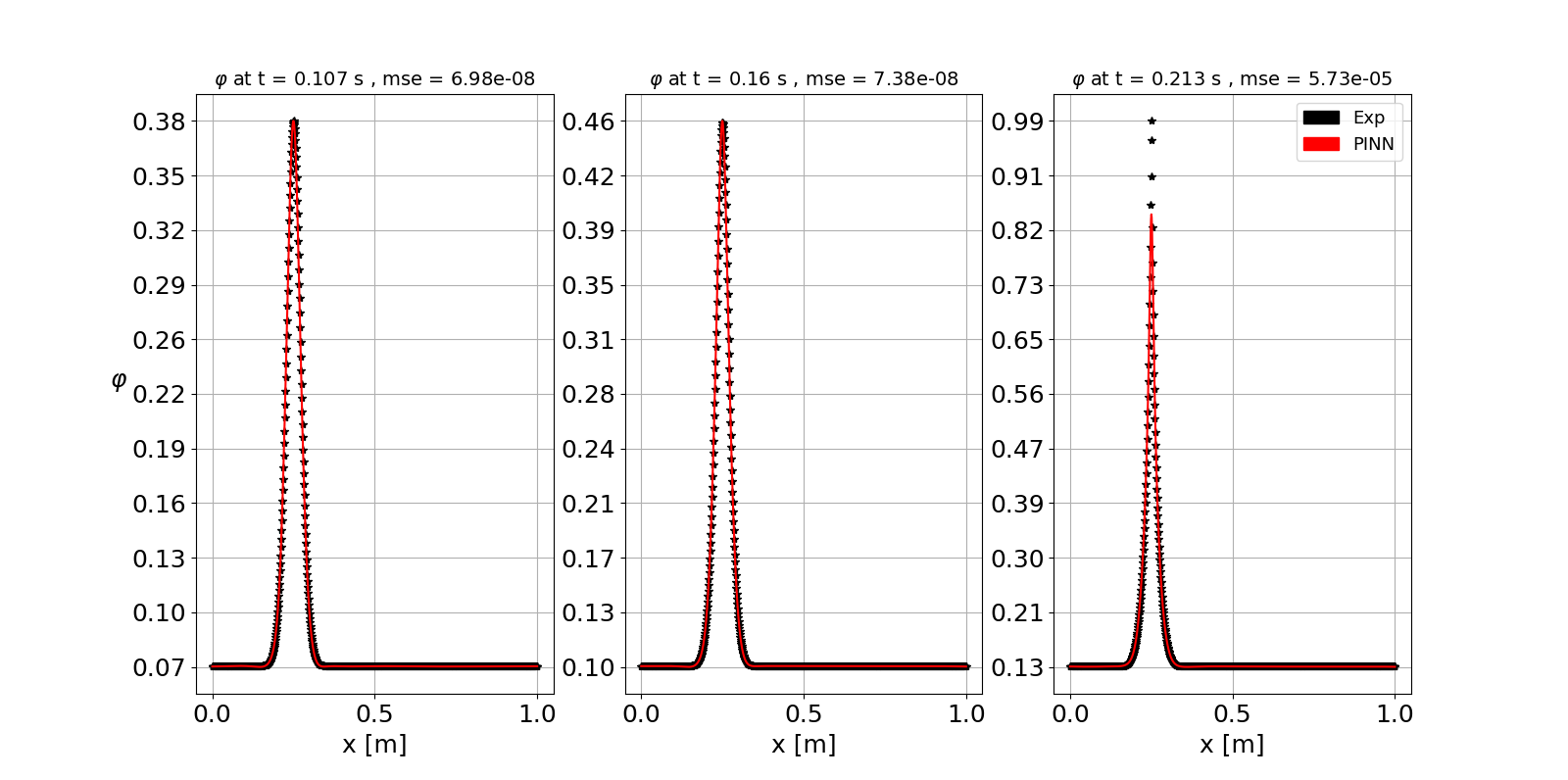}
\caption{Case 2 : Initial condition $\varphi_0^{max}$ $ at $ $x_0=0.25.$}
\label{c2i1}
\end{figure}

\begin{figure}[H]
\centering
\includegraphics[width=0.9\linewidth]{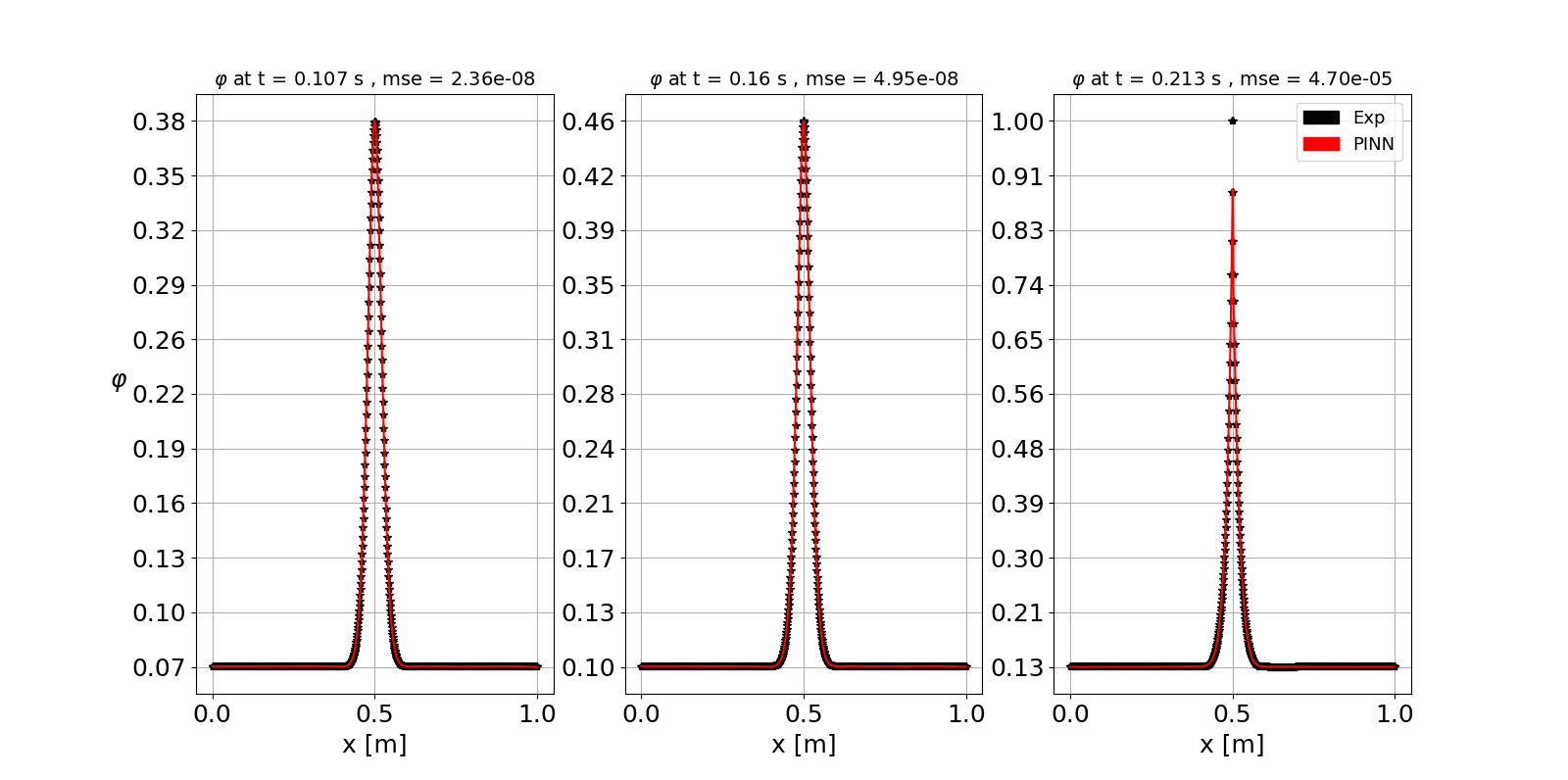}
\caption{Case 2 : Initial condition $\varphi_0^{max}$ $ at $ $x_0=0.5.$}
\label{c2i2}
\end{figure}

\begin{figure}[H]
\centering
\includegraphics[width=0.9\linewidth]{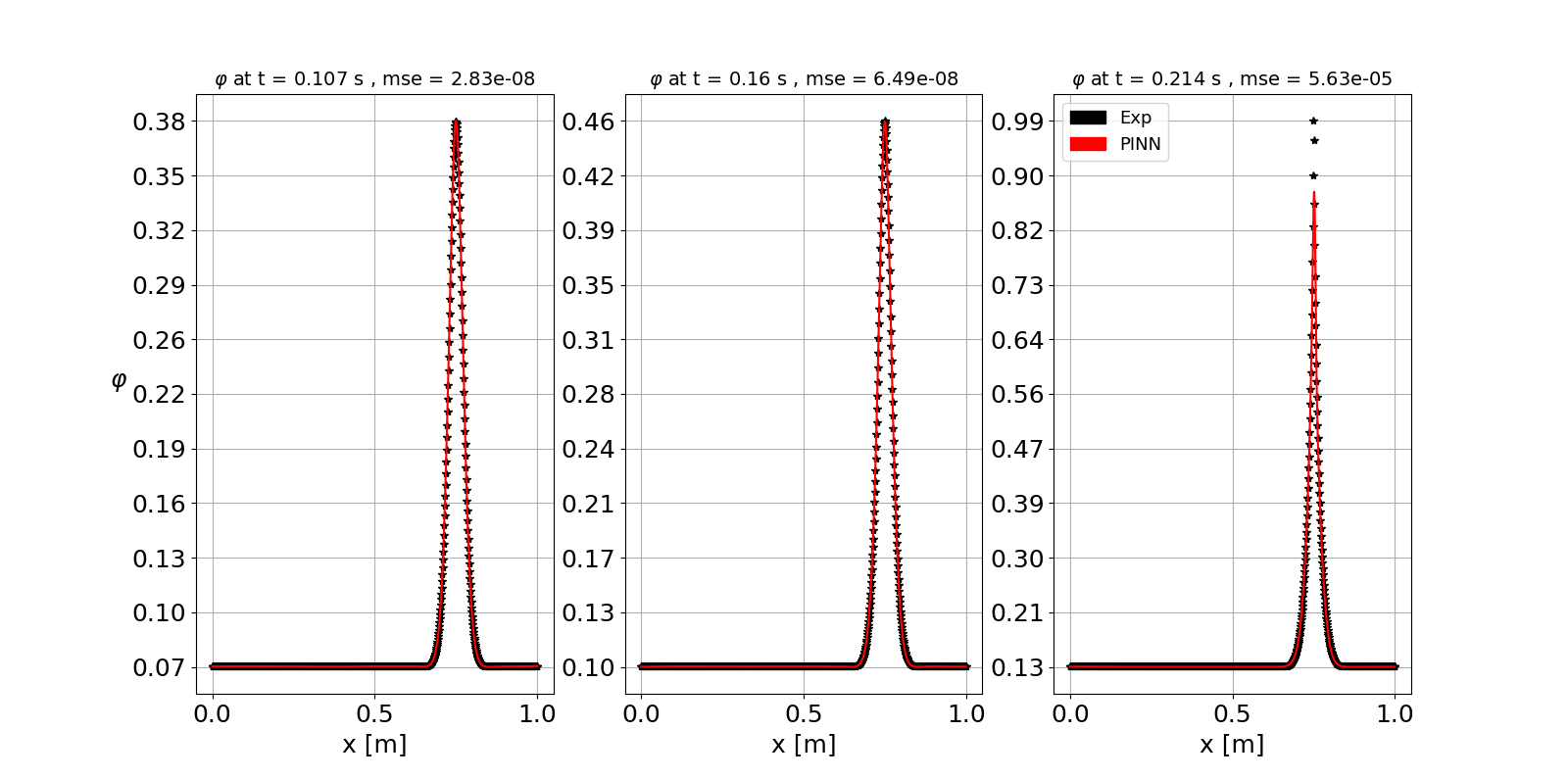}
\caption{Case 2 : Initial condition $\varphi_0^{max}$ $ at $ $x_0=0.75.$}
\label{c2i3}
\end{figure}

In \cref{c2i1,c2i2,,c2i3}, we perceive the same smoothing effect for the last time step of damage evolution and only slight differences in previous steps for all the initial conditions considered.


\subsubsection{Case 3}

In this case, the bar is fixed at the end $x=0$, subjected to a displacement $u_L = \SI{1e-3}{\meter}$  at the end $x=L$ and has a constant distributed load $f_r = \SI{150} {\kilo\newton\per\meter}$. For this and the next case, the strain is calculated from the solutions of the displacement equation
\vspace{5mm}
\begin{equation}
\rho \frac{\partial^2 u }{\partial t^2 }=\frac{\partial }{\partial x }\Bigl((1-\varphi)^{2} E\frac{\partial u}{\partial x } + \Bigr) + f_r(x).\label{Eq36}
\vspace{5mm}
\end{equation} The initial conditions tested  for damage are applied using a
Gaussian distribution centered at $x = 0.25$ , $x = 0.5$ and $x = 0.75$ with  $\varphi_0 = 0.25$.

\noindent We used the same hyperparameters of the last case and the material parameters were estimated with an acceptable percentage error for the initial conditions centered at  $x = 0.5$ and  $x=0.75$. The initial condition centered at  $x = 0.25$ required a larger amount of neurons as showed in \cref{table:14}. 

\begin{table}[H]
\centering
\caption{Hyperparameters of the PINN for case 3.}
\label{table:14}
\vspace{6pt}
\begin{tabular}{ccccccc}
\hline
Layers and neurons& Samples & L-BFGS  & Batch Size & Adam + L-BFGS &  L-BFGS  \\ \hline
[2,20,20,20,20,20,1]   &  8000 & 1000 & 2000 & 10000-80(500) & 20000 \\ \hline
\end{tabular}\end{table}

The percentage error of the parameter identification is shown in \cref{table:15}. In this case, we had a larger error in $\lambda_3$ and a smaller error in $\lambda_2$ for all the initial conditions proposed. 
Like in the previous case, we employed more neurons for one of the  initial conditions considered. This was necessary to have an acceptable percentage error for the material parameters. 

\vspace{7pt}
\begin{table}[H]
\centering
\caption{Percentage errors in the identification for case 3.}
\label{table:15}
\vspace{6pt}
\begin{tabular}{ccllc}
\hline
              &  Parameter & Label value & Estimated value & Percentage error\\ \hline
\multirow{3}{6em}{$\varphi_0^{max}$ $ at $ $x_0=0.25$}
& $\lambda_1$
&  $\num{6.00e3}$ & $\hphantom{xy} \num{5.76e3}$  & $4.03 \%$ \\
& $\lambda_2$
&  $\num{1.70e5} $ & $\hphantom{xy}  \num{1.75e5}$  & $2.77 \%$ \\ 
& $\lambda_3$
&  $\num{2.00e-2} $ & $\hphantom{xy} \num{2.14e-2}$  & $7.06 \%$ \\ \hline

\multirow{3}{6em}{$\varphi_0^{max}$ $ at $ $x_0=0.5$}
& $\lambda_1$
&  $\num{6.00e3}$ & $\hphantom{xy} \num{6.04e3}$  & $0.71 \%$ \\ 
& $\lambda_2$
&  $\num{1.70e5} $ & $\hphantom{xy}  \num{1.72e5}$  & $0.97 \%$ \\ 
& $\lambda_3$
&  $\num{2.00e-2} $ & $\hphantom{xy} \num{2.08e-2}$  & $4.07\%$ \\ \hline

\multirow{3}{6em}{$\varphi_0^{max}$ $ at $ $x_0=0.75$}
& $\lambda_1$
&  $\num{6.00e3}$ & $\hphantom{xy} \num{6.22e3}$  & $3.69 \%$ \\
& $\lambda_2$
&  $\num{1.70e5} $ & $\hphantom{xy}  \num{1.72e5}$  & $1.14 \%$ \\ 
& $\lambda_3$
&  $\num{2.00e-2} $ & $\hphantom{xy} \num{2.17e-2}$  & $8.48 \%$ \\ \hline
\end{tabular}\end{table}

\begin{figure}[H]
\centering
\includegraphics[width=0.9\linewidth]{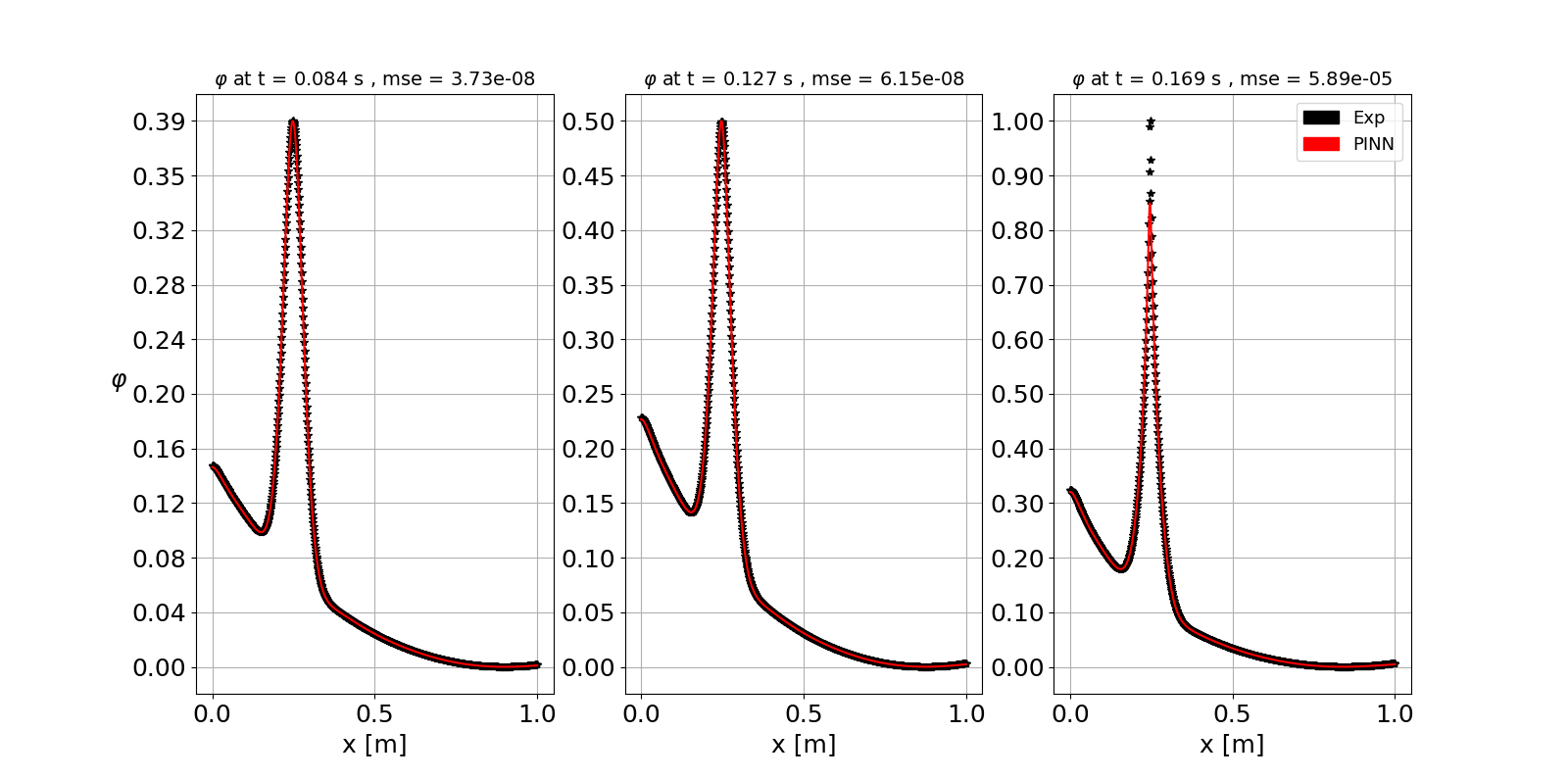}
\caption{Case 3 : Initial condition $\varphi_0^{max}$ $ at $ $x_0=0.25.$}
\label{c3i1}
\end{figure}

\begin{figure}[H]
\centering
\includegraphics[width=0.9\linewidth]{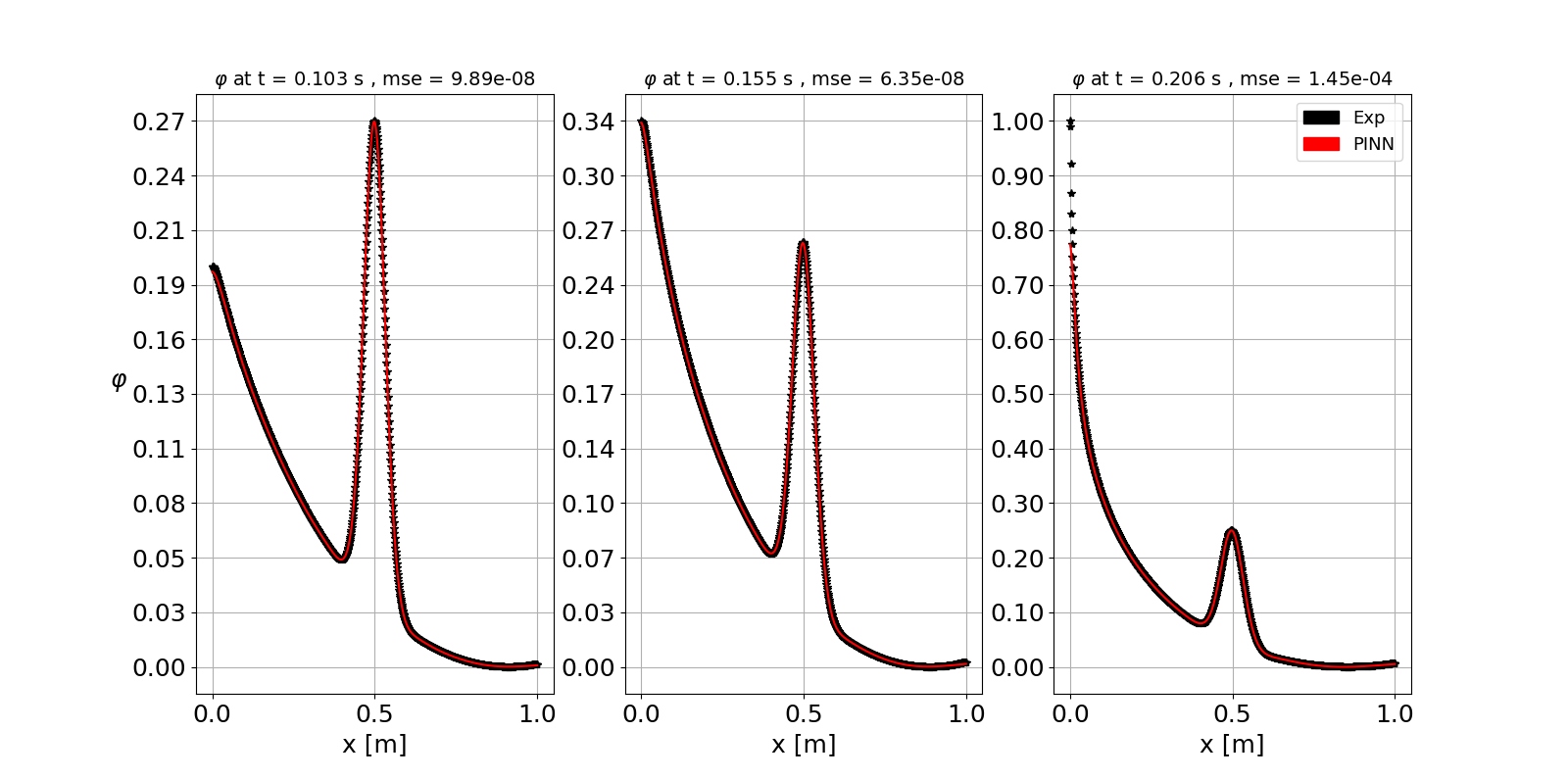}
\caption{Case 3 : Initial condition $\varphi_0^{max}$ $ at $ $x_0=0.5.$}
\label{c3i2}
\end{figure}

\begin{figure}[H]
\centering
\includegraphics[width=0.9\linewidth]{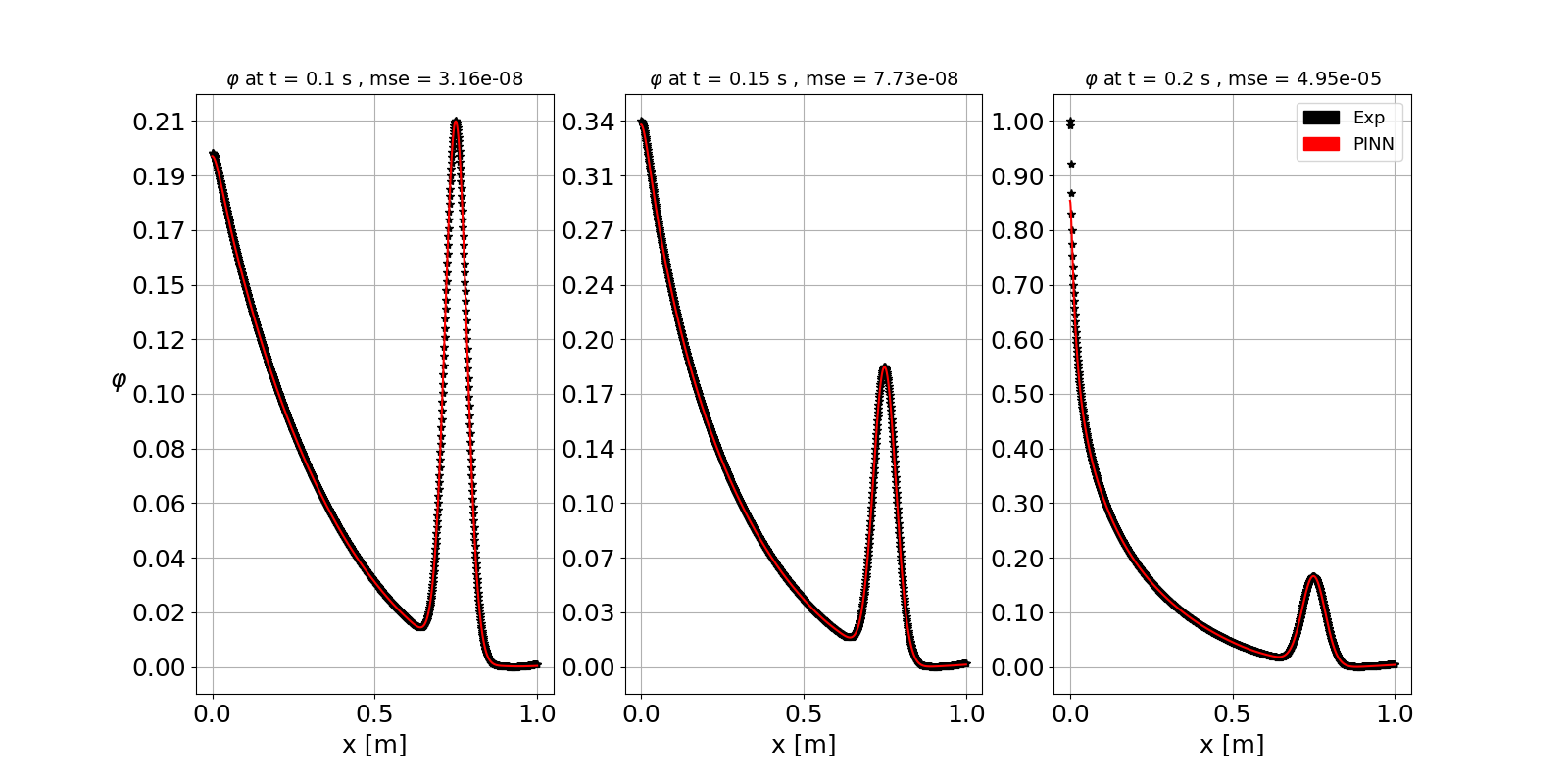}
\caption{Case 3 : Initial condition $\varphi_0^{max}$ $ at $ $x_0=0.75.$}
\label{c3i3}
\end{figure}

In \cref{c3i1,c3i2,c3i3}, it can be seen  that although there are significant variations in the damage behavior depending on the initial conditions proposed, the neural network recovers the pseudo-experimental data in each of them. In contrast with previous cases, the differences of the neural network in the last time steps for initial conditions centered around $x=0.5$ and $x=0.75$ can not be directly attributed to an smoothing effect. However, we see how the values of damage change rapidly in an small $\Delta x$ which is something that also happened before and after the peak of the previous cases. 


\subsection{Case 4}

In this case, the bar is fixed at the end $x=0$, subjected to a displacement $u_L = \SI{1e-3}{\meter}$  at the end $x=L$ and has a  distributed load $f_r = 150\sin \Bigl(\frac{\pi x}{2}\Bigr)$ \SI{}{\kilo\newton\per\meter}. The three initial conditions tested  for damage are applied using a
Gaussian distribution centered at $x = 0.25$ , $x = 0.5$ and $x = 0.75$ with  $\varphi_0 = 0.35$.

For this case, we use a neural network with 6 hidden layers of 14 neurons for the initial conditions centered at  $x = 0.25$ and  $x=0.75$ and an architecture with 5 hidden layers and the same number of neurons for the initial condition centered at  $x = 0.5$. All the configurations of the optimization stages are the same of the previous cases.

The percentage errors of the parameter identification are given in \cref{case44}. In this case, we had a larger error in $\lambda_3$  for all the initial conditions proposed. 
Similarly to the previous cases (2 and 3), we needed a slightly different configuration for two of the initial conditions considered. 

\vspace{7pt}
\begin{table}[!ht]
\centering
\caption{Percentage error in the identification for case 4.}
\label{case44}
\vspace{6pt}
\begin{tabular}{ccllc}
\hline
              &  Parameter & Label value & Estimated value & Percentage error\\ \hline
              
\multirow{3}{6em}{$\varphi_0^{max}$ $ at $ $x_0=0.25$}
& $\lambda_1$
&  $\num{6.00e3}$ & $ \hphantom{xy} \num{5.57e3}$  & $7.16 \%$ \\
& $\lambda_2$
&  $\num{1.70e5} $ & $\hphantom{xy}  \num{1.82e5}$  & $7.11 \%$ \\ 
& $\lambda_3$
&  $\num{2.00e-2} $ & $\hphantom{xy} \num{2.22e-2}$  & $11.25\%$ \\ \hline

\multirow{3}{6em}{$\varphi_0^{max}$ $ at $ $x_0=0.5$}
& $\lambda_1$
&  $\num{6.00e3}$ & $\hphantom{xy} \num{6.02e3}$  & $0.33 \%$ \\
& $\lambda_2$
&  $\num{1.70e5} $ & $\hphantom{xy}  \num{1.71e5}$  & $0.68 \%$ \\ 
& $\lambda_3$
&  $\num{2.00e-2} $ & $\hphantom{xy} \num{2.04e-2}$  & $1.81\%$ \\ \hline

\multirow{3}{6em}{$\varphi_0^{max}$ $ at $ $x_0=0.75$}
& $\lambda_1$
&  $\num{6.00e3}$ & $\hphantom{xy} \num{6.16e3}$  & $2.65 \%$ \\
& $\lambda_2$
&  $\num{1.70e5} $ & $\hphantom{xy}  \num{1.71e5}$  & $0.66 \%$ \\ 
& $\lambda_3$
&  $\num{2.00e-2} $ & $\hphantom{xy} \num{2.09e-2}$  & $4.57 \%$ \\ \hline
\end{tabular}\end{table}

\begin{figure}[H]
\centering
\includegraphics[width=0.9\linewidth]{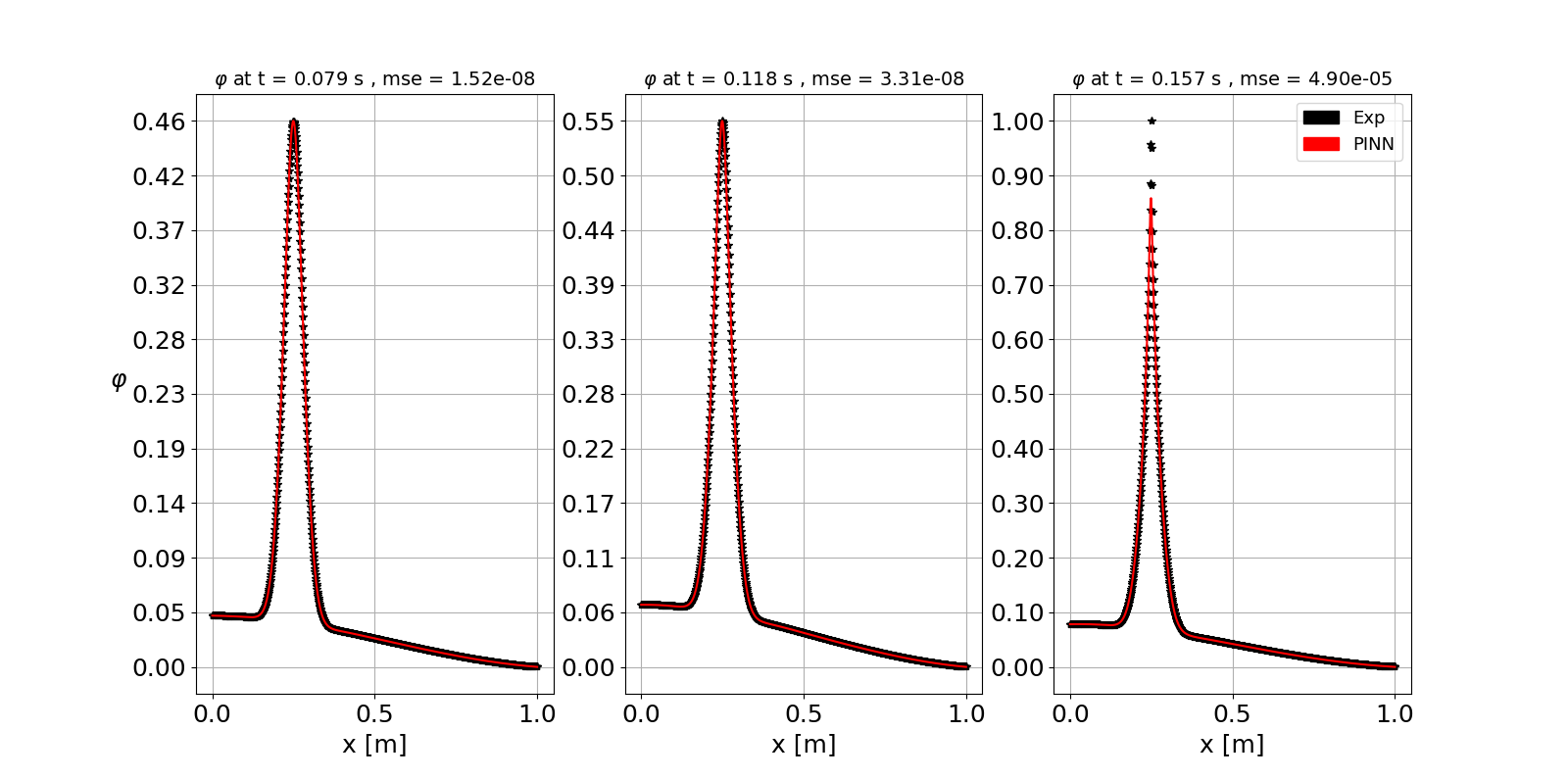}
\caption{Case 4 : Initial condition $\varphi_0^{max}$ $ at $ $x_0=0.25.$}
\label{c4i1}
\end{figure}
\begin{figure}[H]
\centering
\includegraphics[width=0.9\linewidth]{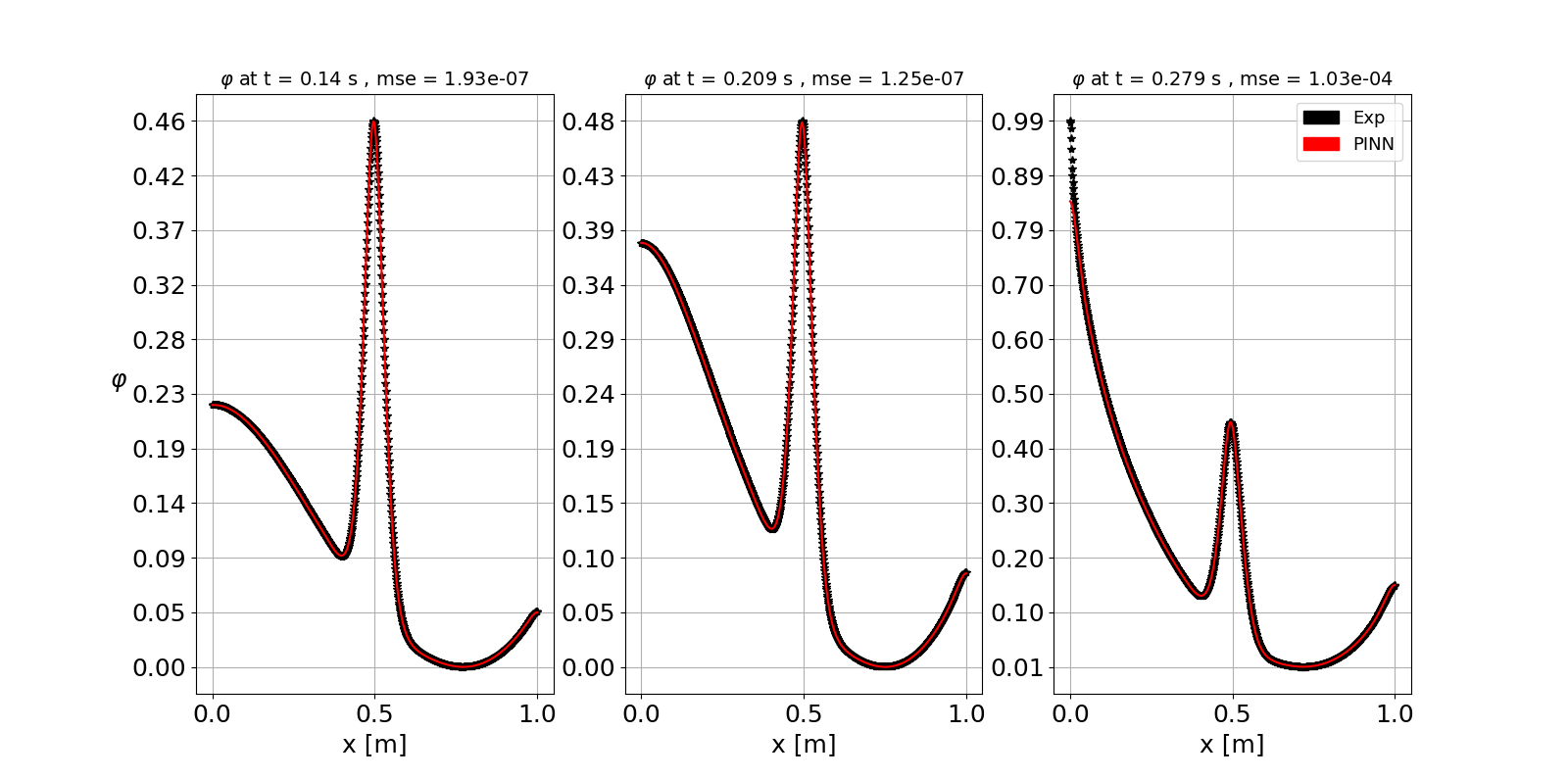}
\caption{Case 4 : Initial condition $\varphi_0^{max}$ $ at $ $x_0=0.5.$}
\label{c4i2}
\end{figure}
\begin{figure}[H]
\centering
\includegraphics[width=0.9\linewidth]{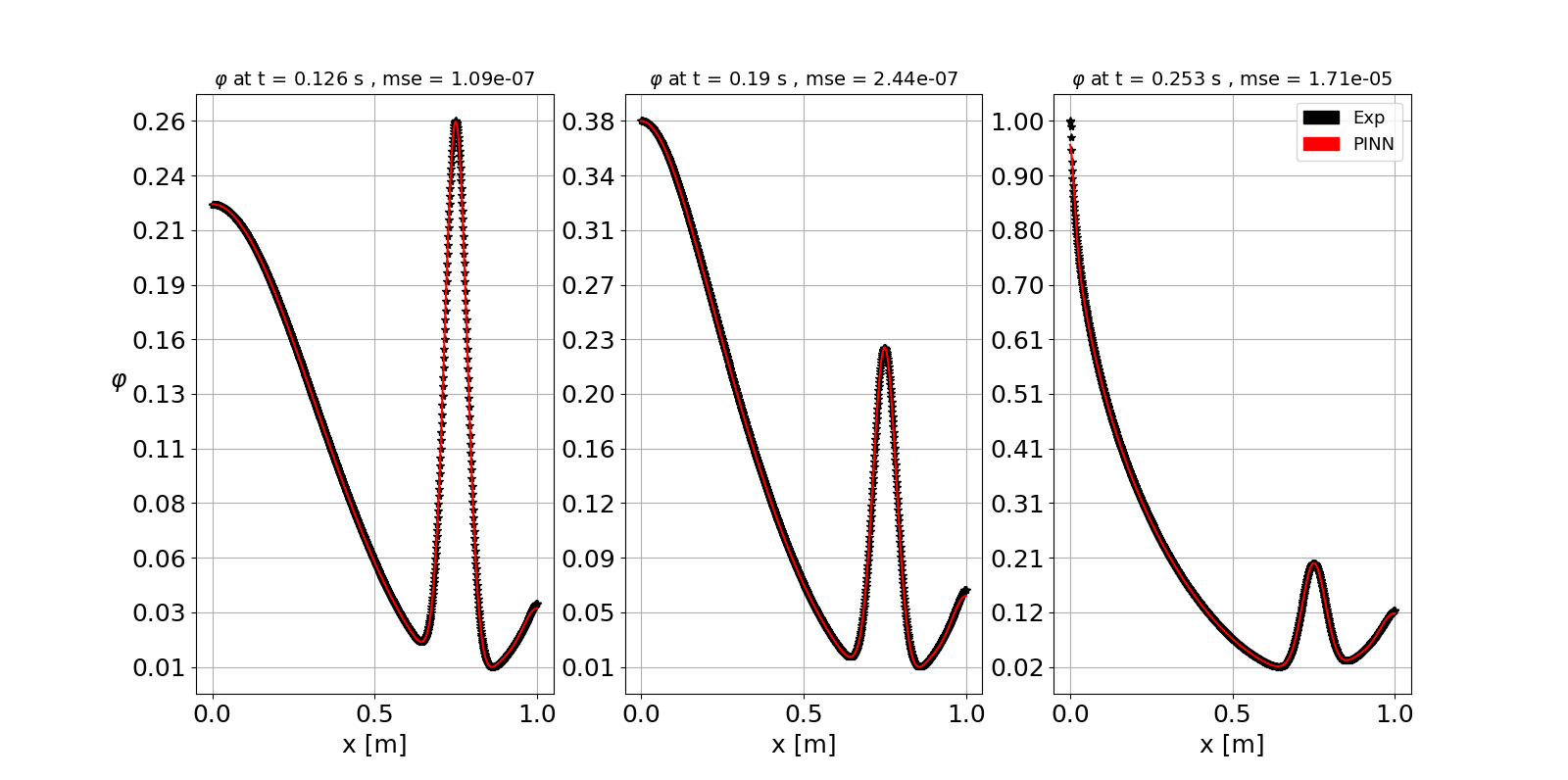}
\caption{Case 4 : Initial condition $\varphi_0^{max}$ $ at $ $x_0=0.75.$}
\label{c4i3}
\end{figure}

In \cref{c4i1,c4i2,c4i3}, we see that, as in the previous case, there are significant variations in the damage evolution depending on the initial conditions proposed. Nevertheless, the neural network recovered well the pseudo-experimental data in each of them. It can be noticed that the approximation in the last step is better for the initial condition centered around $x=0.75$ than $x=0.25$. This shows that the hyperparameters selected are better in some cases and depending on the conditions is better to adopt particular considerations for each physical case.

\section{Conclusions}

This work addressed the estimation of parameters in the governing equations of the damage model proposed in \cite{boldrini2016non}. We implemented the identification of three material parameters applying a physics informed neural network and combining some ideas of the two-step method, the principal differential analysis, and the generalized smoothing approach. Initially, the identification strategy was tested using only the evolution of the damage and afterward, it was included an additional input in the neural network model with information from the solution of the displacement equation. We examined the robustness of the method in the presence of noisy training data and also their generalization capabilities in different physical cases.

In \cref{pid_sec} we introduced the formulation of the material identification as an optimization problem and described some popular techniques that have been used over the years. There are different levels of classification for the methods to solve this type of inverse problem, but it seems that exists a trend towards the adoption of function approximators to fit the models using Bayesian inference. These choices are popular because they work well in the presence of noise and also can introduce prior information in the solution to the problem. In this work, we used a neural network as function approximator and a deterministic approach to estimate the parameters.

We adopted the physics informed methodology as the base for our neural network model because it has been proved in different identification problems with good results. However, we decided to make some modifications in the formulation of the optimization process because we observed that when the term that fits the observations and the residue of the governing equation of the model have the same importance, the neural network demands a higher number of neurons and layers. We also now from the work in \cite{ramsay_parameter_2007}, that the simultaneous search for the parameters of the neural network, i.e., nuisance parameters, can complicate the estimation of the material parameters, i.e., structural parameters. In order to avoid this and local minima trouble, we proposed three stages of optimization. First, we optimize the collocation loss which means that in this stage only the parameters of the network are refined. In the second stage, we alternate between a simultaneous search of all the parameters using a gradient descent method, and the optimization of only the neural network parameters using an L-BFGS algorithm. Lastly, we complemented the estimation of the parameters implementing a simultaneous search with a high execution limit.

The equations of the model for the hypothesis presented in \cref{damage model} give the evolution of the displacement and damage and were applied in a simple physical case. We used a bar with the left extreme fixed and diverse boundary conditions for the right end. With the purpose of exploring different approaches, we simplified the identification problem considering only the damage equation with a constant strain in the bar. We tested the robustness of the implementation using various levels of noise in the training data. Although the mean errors for the different levels of noise were larger than those of the clean data, the quantitative results were within an acceptable range considering that we kept the same hyperparameters tuned for the original (clean) training data. In addition, we presented the results of the estimation using a method that constrains the optimization using the numerical solution of the differential equation. Despite that the results for the clean data were superior, we found that the mean error in the presence of noise was higher for the constrained method. In this case, the neural network methodology had better performance when dealing with noisy data which is a property desired to work with experimental observations.

Finally, we proposed four physical cases to evaluate the generalization capabilities of the strategy proposed. These cases had different boundary conditions, initial conditions and some included distributed loads. We used a random search to tune one of the configurations in each case and only introduced minor changes in the number of neurons and layers to maintain the error controlled when it was necessary. In general, just slight modifications were needed and the quantitative and qualitative results were satisfactory.

\bibliographystyle{unsrt}  
\bibliography{references}  






\end{document}